\documentclass[a4paper,usenatbib]{mnras}

\usepackage{newtxtext,newtxmath}
\usepackage{amsmath}
\usepackage[T1]{fontenc}
\usepackage{ae,aecompl}
\usepackage[dvipsnames]{xcolor}

\usepackage{graphicx}	% Including figure files
\graphicspath{{./img/}}

\usepackage{amsmath}	% Advanced maths commands

\usepackage{subcaption}
\usepackage{gensymb}
\usepackage{makecell}
\usepackage{tablefootnote}

\newcommand{\bs}{\boldsymbol}

\newcommand\psj{PSJ}

\defcitealias{Rucska21}{Paper I}

\title[Planetesimal formation with multiple grains]{Planetesimal formation via the streaming instability with multiple grain sizes}

\author[J. J. Rucska and J. W. Wadsley]{
J. J. Rucska,$^{1}$
J. W. Wadsley$^{1}$\thanks{E-mail: wadsley@mcmaster.ca}
\\
$^{1}$Department of Physics and Astronomy, McMaster University, Hamilton, L8S 4M1, Canada
}

\date{Accepted XXX. Received YYY; in original form ZZZ}

\pubyear{2020}

\begin{document}
\label{firstpage}
\pagerange{\pageref{firstpage}--\pageref{lastpage}}
\maketitle

\begin{abstract}
Kilometre-sized planetesimals form from pebbles of a range of sizes. We present the first simulations of the streaming instability that begin with a realistic, peaked size distribution, as expected from grain growth predictions. Our 3D numerical simulations directly form planetesimals via the gravitational collapse of pebble clouds. Models with multiple grain sizes show spatially distinct dust populations. The smallest grains in the size distribution do not participate in the formation of filaments or the planetesimals that are formed by the remaining $\sim$80\% of the dust mass. This implies a size cutoff for pebbles incorporated into asteroids and comets. Disc observations cannot resolve this dust clumping. However, we show that clumping, combined with optical depth effects, can cause significant underestimates of the dust mass, with 20\%-80\% more dust being present even at moderate optical depths if the streaming instability is active.

\end{abstract}


%%%%%%%%%%%%%%%%%%%%%%%%%%%%%%%%%%%%%%%%%%%%%%%%%%

%%%%%%%%%%%%%%%%% BODY OF PAPER %%%%%%%%%%%%%%%%%%

% Introduction
\section{Introduction}

In the process of planet formation, planetary embryos grow from the collisions of many millions of 1 km to 100 km sized planetesimals. In turn, planetesimals are born from millimetre-centimetre sized pebbles in protoplanetary discs. However, the formation of planetesimals cannot occur through simple pathways such as the collisional coagulation of progressively larger objects. It is well understood that collisions between objects in the range of 1 cm to 1 m in protoplaneteary disc environments are predominantly destructive, resulting in smaller remnants from the original bodies in the collision \citep{Zsom10,Guttler10,Windmark12}. Further, all solid objects orbiting in protoplanetary discs experience a headwind as they orbit through the gaseous component of the disc. At $\sim$1 m sizes, this process is maximally efficient, and causes the rapid orbital decay of these objects, sending them into the central star on timescales on the order of a few hundred years \citep{Weiden77}.

Hence, planet formation requires a mechanism that is capable of rapidly forming planetesimals directly from cm sized pebbles. A leading candidate for this process is known as the streaming instability (SI), first studied by \citet{YG05} (see also: \citealt{YJ07,JY07}). The SI is a specific example of a broader family of resonant drag instabilities that exist when the aerodynamic drag timescale becomes resonant with another dynamical timescale in the disc \citep{SquireHop18,SquireHop20}. In the saturated, non-linear phase of the instability, the SI is capable of producing strong, localized overdensities of clouds of pebble-sized dust that can then gravitationally collapse into planetesimals \citep{Johansenetal07}, thus directly overcoming the aforementioned growth barriers.

Since the seminal work from \citet{Johansenetal07}, over a decade of research has explored planetesimal formation via the SI with high resolution 3D hydrodynamic simulations \citep{Johansen09, Johansen12, Johansen15, Simon16, Simon17, Schafer17, Abod19, Li19, Nesvorny19, Nesvorny21, Gole20, Rucska21, Carrera21, Carrera22a, Carrera22b}. The streaming instability has proven to be a robust mechanism for forming planetesimals, so long as the local protoplanetary disc region meets the prerequisite conditions of enhanced dust mass concentration (i.e. supersolar) and sufficiently large dust grains \citep{Carrera15,Yang17,Li21}. Protoplanetary discs observed with the Atacama Large Millimeter/submillimeter Array (ALMA) and the Very Large Telescope (VLT)/SPHERE and Subaru/HiCIAO have shown features with concentrated dust mass, such as rings \citep[e.g.][]{Dull18-DSVI, Macias19, Muto12, Avenhaus18, Long18}, non-axisymmetric bumps \citep[e.g.][]{vdM13, vdM15, Cazzoletti18, vdM21-asym} and spiral structure \citep[e.g.][]{Benisty15, Perez16, Benisty17}. Some rings may have sufficiently high dust concentrations to initiate planetesimal formation via the SI \citep{Stammler19, Mauco21}. Indeed, \citet{Carrera21}, \citet{Carrera22a} and \citet{XuBai22a,XuBai22b} show that persistent radial gas pressure maxima, which likely play a role in the formation of the observed large-scale rings \citep{Whipple72}, can sufficiently concentrate dust to trigger planetesimal formation via the SI.

Observations of minor solar system bodies support the idea that these objects may have been formed via the SI or a similar process. Asteroids are commonly described as ``rubble piles'': gravitationally bound conglomerates of smaller pebbles with significant bulk porosity \citep{Walsh18}. Results from the Rosetta mission to comet 67P/Churyumov-Gerasimenko suggest this object likely formed from a cloud of millimetre-sized dust particles \citep{Blum17, Fulle17}. Further, results from the New Horizons flyby of the Kuiper belt object (486958) Arrokoth suggest that this object--a contact binary with two distinct lobes--is likely a result of the slow decay of a binary orbit, where the two progenitor objects formed via the gravitational collapse of a pebble cloud \citep{McKinnon20,Grishin20,Marohnic21}. Arrokoth is likely a primitive object which has experienced little to no collisional evolution since formation. The term rubble pile, as commonly used in planet formation, refers to the loose assembly of material and does not require that the object is composed of collisional debris. \citet{Nesvorny19} and \citet{Nesvorny21} also show that the gravitational collapse of dense pebble clouds from SI simulations can produce planetesimal binaries with properties similar to binaries observed in the Kuiper belt. \citet{Kavelaars21} measured the size distribution of objects in the cold classical Kuiper belt and find it is well described by an exponential cut-off at large sizes---a feature predicted by the streaming instability.

The New Horizons mission also observed the craters that cover the 4 billion-year-old surfaces of the Pluto-Charon system, enabling an analysis of the inferred size distribution of the impactors from the early Solar system that produced those craters \citep{Singer19,RobbinsSinger21,Robbins17}. \citet{Singer19} find a deficit of craters at small sizes, for impactors below $\lesssim$ 1-2 km in diameter. Unfortunately, current limits on computational power prevent 3D simulations of the SI from providing any insights on the SI-formed planetesimal size distribution at these small sizes (\citealt{Simon16}; \citealt{Li19}; see Section 4.1 of \citealt{Rucska21} for further discussion).

Observational constraints on planetesimal formation are difficult to acquire, yet there is a general agreement between observational data and predictions from models of planetesimal formation via the SI. To date, the SI remains a leading candidate for the efficient formation of planetesimals, yet there remains open questions regarding this process, such as how the presence of a distribution of dust grain sizes affects outcomes regarding planetesimal formation.

\subsection{Dust grain size distributions in protoplanetary discs}

Observations reveal that protoplanetary discs in nature have at least two distinct dust populations \citep[e.g.][]{Franc23}: $\sim$millimetre-sized pebbles which have settled to the disc mid-plane and are most readily visible via their sub-mm wavelength thermal emission with ALMA \citep[e.g.][]{Andrews16, vdM21-asym, Mauco21}, and $\sim$micron-sized grains suspended vertically in the disc, seen in infrared scattered light \citep[e.g.][]{Muto12,Benisty15,Avenhaus18}. Though these components occur in spatially distinct regions in the disc, they are likely linked, as grain growth theory shows that pebbles can readily grow via coagulation from the micron-size grains that the disc inherits from the interstellar medium (\citealt{Birnstiel11,Birnstiel15}, see \citealt{Birnstieletal16}, for a review). Once the grain growth/fragmentation process reaches equilibrium, the predicted outcome from the widely-used \citet{Birnstiel11} model is a grain size distribution described by multiple power-laws and a distinct peak, so that most of the mass in the distribution is within a factor of two of a specific grain size. 

\subsection{Streaming instability with a distribution of grain sizes}

Until recently, there were few studies of the streaming instability with multiple sizes. \citet{Johansenetal07} included multiple dust species in a subset of their runs, but the focus of their work was the onset of planetesimal formation rather than the behavior of the different grains. \citet{Bai102} modelled discs with a variety of grain size distributions simultaneously in 3D simulations, and explored the influence of these distributions on properties of the non-linear, saturated state of the SI, pre-planetesimal formation.

Recently, there have been multiple studies on how particle size distributions influence the linear growth phase of the SI \citep{Krapp19, Paardekooper20, Paardekooper21, McNally21, ZhuYang21} and the linear and non-linear phase in 2D numerical simulations \citep{Schaffer18, Schaffer21, YangZhu21}. These studies explored linear SI growth rates and the clumping of dust in the non-linear phase for distributions with a wide range of grain sizes. Overall, they conclude that the SI can produce strong dust clumping so long as the local dust-to-gas mass density ratio is large, approaching unity, and that the grain size distribution involves sufficiently large grains (near approximately a centimetre in size). In this paper, we study dust that is well within the strong growth regime, and follow the non-linear phase of the SI all the way to planetesimal formation.

We expand upon prior work by \citet{Bai102} (3D), \citet{Schaffer18, Schaffer21} (2D) and \citet{YangZhu21} (2D). The grain size distributions in these studies are power laws, with exponents similar to the fiducial slope for interstellar grains from \citet{Mathis72}, the so-called ``MRN distribution''. In this study, we sample the grain size distribution of \citet{Birnstiel11}, which is the equilibrium outcome of a grain growth/fragmentation model applicable to the midplane of protoplanetary discs, where planetesimal formation is believed to occur. The \citet{Birnstiel11} distribution deviates from a single power law and includes a peak at large sizes. Thus, in our discretized version of that grain size distribution, the spacing between the representative grain size for each bin is not equal, in linear or logarithmic space, which is unique from prior work on this subject.

We present the first 3D, vertically stratified simulations of the SI with multiple species of dust grains since \citet{Bai102}, and compare the non-linear development of the SI in dust with multiple sizes against data from our prior work which used a single size \citep{Rucska21}. We highlight the differences in the dust surface density distribution between multi-size and single-size models, along with a novel analysis that reveals the observational consequences of the strong dust clumping seen in our runs, and explore how grains of different sizes participate in planetesimal formation.

Our paper is organized as follows. In Section~\ref{sec:methods} we present our methods and choice of parameters and a discussion about the dust grain size distribution we model. Section~\ref{sec:res_surfden} focuses on the different dust surface density distributions between our multi-size model and prior work with single grain sizes, and the observational consequences of these differences. Section~\ref{sec:res_clumps} focuses on how the different grain sizes participate in the non-linear filament and planetesimal formation process. In Section~\ref{sec:discsumm3} we summarize our key results and discuss how this paper influences the current understanding of planetesimal formation via the SI. 

% Methods
\section{Methods}
\label{sec:methods}

We model a local portion of a near-Keplerian protoplanetary disc. We study the dynamics of a gas phase aerodynamically coupled to a dust/solids phase. The specifics of our numerical and hydrodynamic set-up are nearly identical to those described in \citet{Rucska21}, so we briefly summarize those methods here and refer a reader interested in a more detailed discussion to that paper.

We use the shearing sheet approximation of \citet{GLB65I} to track the local dynamics of a Cartesian frame co-rotating at the Keplerian orbital velocity. We employ the \textsc{Athena} hydrodynamics code \citep{StoneAth08,StoneGard09} with the solids particle module \citep{Bai101} to numerically evolve the protoplanetary disc system. Vertically, the box is centred on the disc midplane ($z=0$), and the co-rotating frame of reference leads to an imposed background velocity in the azimuthal ($y$) direction described by $(q\Omega x)\bs{\hat{y}}$. Here $x$ is the radial co-ordinate in the co-rotating frame, with $x=0$ being the radial centre of the box, and $q$ is the power-law index of the angular velocity with radial position in the disc, $\Omega \propto r^{-q}$, so that in Keplerian discs $q=3/2$.

The equations that describe the dynamics of the gas and solids (dust) are
\begin{align}
 {\frac{\partial \rho_g}{\partial t }} & + \nabla \cdot (\rho_g \bs{u}) = 0 , \\
 \label{eq:gasmom}
 {\frac{\partial \rho_g \bs{u} }{\partial t}} & + 
\nabla \cdot (\rho_g \bs{u}\bs{u}) =  -\nabla P_g \nonumber \\ & + \rho_g \Bigg[-2\bs{\Omega} \times \bs{u} + 2q\,\Omega^2 x\, \hat{\bs{x}}
 -\Omega^2 z\, \hat{\bs{z}} + \mu \frac{\overline{\bs{v}} - \bs{u}}{t_{\text{stop}}} \Bigg],  \\
 \label{eq:dustmomprime3}
\frac{d \bs{v}_{i}' }{dt} & = 2(v_{iy}' - \eta v_K) \Omega \hat{\bs{x}} - (2 - q)v_{ix}'\Omega \hat{\bs{y}} - \Omega^2 z \hat{\bs{z}} - \frac{\bs{v}_i' - \bs{u}'}{t_{\text{stop}}} + \bs{F}_g,
\end{align}
where $\rho_g$ is the gas mass density, $P_g$ is the gas pressure, $\bs{u}$, is the velocity of the gas, $\bs{v}_i'$ is the velocity of an individual dust particle in the frame of the background shear flow, and $\overline{\bs{v}}$ is the mass-weighted average velocity of the dust in a gas cell. The gas equation of state is isothermal, $P_g = \rho_gc_s^2$, where $c_s$ is the sound speed. The quantity $\mu \equiv \rho_d/\rho_g$ is the local ratio of dust to gas mass density, and $\eta$ controls the strength of the radially inward drag force on the dust, which is related to the steepness of the radial gas pressure gradient (see Section~\ref{sec:physparam}). The quantity $t_{\text{stop}}$ is the time-scale for the exchange of momentum between the dust and gas phase, which depends on a local gas quantities such as density and temperature, and, crucially, the physical size of the dust grains. We discuss this parameter in more detail in Section~\ref{sec:gsd} as it is central to the context for this paper.

For the numerical algorithms, as in \citet{Rucska21} we use the standard \textsc{Athena} options for the Reimann solver (HLLC), hydrodynamics integrator (corner transport upwind) and a semi-implicit integrator for the dust momentum equations with a triangular-shaped cloud scheme to interpolate the dust particle properties with the simulation grid. In equation~\ref{eq:dustmomprime3}, the background shear flow has been subtracted from the dust and gas velocities. Separating the advection of the shear velocity from local deviations leads to a more efficient and accurate numerical integration \citep{Masset00, Johnson08}. We use the shearing box boundary conditions, which are periodic in the azimuthal ($y$) and the vertical directions ($z$) and shear periodic in the radial ($x$) direction \citep{Hawley95, StoneGard10}.

The term $\bs{F}_g$ in equation~\ref{eq:dustmomprime3} represents the gravitational acceleration. Not all prior work on high-resolution studies of the non-linear SI includes the effects of the dust density field self-gravity, but since our study is in part focused on the properties of planetesimals, it is included here. Self-gravity enables the collapse of dense dust material into gravitationally bound objects (i.e. planetesimals). Following \citet{Simon16}, based on an implementation described and tested in \citet{Rucska21}, this acceleration is computed via the gradient in the gravitational potential from the dust density field, and this potential is computed from the solution to Poisson's equation,
\begin{align}
\bs{F}_g & = -\nabla \Phi_d, \\
 \label{eq:poisson}
\nabla^2 \Phi_d & = 4\pi G \rho_d.
\end{align}
Here $G$ is the gravitational constant. Our parameterization of this constant is discussed further in the next section (equation~\ref{eq:Gtilde}). Note, we neglect the gravitational influence of the gas, since the density perturbations in the gas in these kinds of local protoplanetary disc models are very small \citep{Li18}. We use the fast Fourier transform Poisson solver available in \textsc{Athena} \citep{KimOst17} to solve equation~\ref{eq:poisson}, shear periodic horizontal boundary conditions and vaccuum boundary conditions vertically.

\subsection{Physical and numerical parameters, initial conditions}
\label{sec:physparam}

\begin{table}
 \caption{Simulation parameters.}
 \label{tab:sims3}
 \begin{tabular}{lccc}
  \hline
  Run names & 
  \multicolumn{3}{c}{$\tau_s$ - grain stopping time(s)} \\
  
  \hline  
  
  \makecell[l]{\texttt{S0,S1,}\\\texttt{S2,S3}\\(\texttt{S})} & 
  \multicolumn{3}{c}{0.314} \\
  
  \rule{0pt}{6.5ex}\noindent

  \makecell[l]{\texttt{M6-0,M6-1,M6-2,}\\\texttt{M6-3,M6-4}\\(\texttt{M6})} & 
  \multicolumn{3}{c}{0.036, 0.191, 0.270, 0.314, 0.353, 0.412} \\

  \rule{0pt}{5ex}\noindent
  
  \texttt{M12} & 
  \multicolumn{3}{c}{\makecell{0.021, 0.113, 0.170, 0.218, 0.256, 0.284,\\ 0.305, 0.324, 0.342, 0.363, 0.390, 0.437}} \\
  
  \rule{0pt}{6.5ex}\noindent
  
  \texttt{M18} & 
  \multicolumn{3}{c}{ \makecell{0.016, 0.083, 0.125, 0.162, 0.196, 0.226, \\ 0.251, 0.272, 0.288, 0.302, 0.314, 0.327,\\ 0.339, 0.352, 0.367, 0.385, 0.408, 0.450}} \\
  
  \hline
  \hline
  
  \multicolumn{2}{c}{Domain Size} & 
  \multicolumn{2}{c}{Grid Resolution} \\
  
  \multicolumn{2}{c}{$(L_x \times L_y \times L_z)/H_g$} & 
  \multicolumn{2}{c}{$N_{\text{cell}} = N_x \times N_y \times N_z$} \\
  
  \multicolumn{2}{c}{0.2 $\times$ 0.2 $\times$ 0.2} & 
  \multicolumn{2}{c}{120 $\times$ 120 $\times$ 120} \\
  
  \hline
  \hline
  \rule{0pt}{3ex}\noindent
  
  ${N_{\text{par}}/(N_{\text{species}} \times N_{\text{cell}})}$ &
  $Z$ & $\widetilde{G}$ & $\Pi$\\
  
  \multicolumn{1}{c}{1} & 0.02 & 0.05 & 0.05\\
  
  \hline 
  \rule{0pt}{3ex}\noindent
 \end{tabular}
 %$^{*}$Each grain species individually was given this particle resolution.
\end{table}

In this section we discuss the physical parameters that influence the dynamics of our local protoplanetary disc system. In this study we choose identical or very similar values to previous work on these systems \citep{Simon16,Schafer17,Johansen12,Li18,Gole20,Rucska21}. These parameters and our choices are summarized in Table~\ref{tab:sims3} and briefly discussed in this section, with a more in-depth discussion of the stopping time parameter $\tau_s$ in Section~\ref{sec:gsd}.

The total mass of the dust particles is controlled by the ratio of dust mass surface density to the gas surface density
\begin{equation}
Z = \frac{\Sigma_d}{\Sigma_g},
\end{equation}
and we choose $Z=0.02$, which is a slightly supersolar metal mass ratio. Note that our simulation domains model only a fraction of the vertical gas scale height while capturing the full dust scale height. Thus the effective surface density mass ratio within the simulation domain is higher than 0.02. Following the discussion from Section 2.4 of \citet{Rucska21}, the ratio of total dust mass to total gas mass within the full simulation domain is approximately 0.25.

Within \textsc{Athena}, the effects of the radial gas pressure gradient, which exerts a radially outward force on the gas, is represented by a radially inward force on the dust instead: the $-2\eta v_k \Omega$ term in  equation~\ref{eq:dustmomprime3}  \citep{Bai101}. This force is parameterized via $\eta$,
\begin{equation}
\eta = n\frac{c_s^2}{v_K^2},
\end{equation}
where $n$ is the pressure power law index, $P_g \propto r^{-n}$, the local Keplerian speed is $v_K$, and the isothermal sound speed is $c_s$. Thus, our localized model tracks the gas in a slightly sub-Keplerian, pressure-supported reference frame, which shifts the azimuthal component of the dust and gas velocities by $\eta v_K$. We then subtract this shift from our data to conduct analysis in this shifted frame.

As with other work, in our simulations $\eta$ is ultimately controlled by a similar parameter
\begin{equation}
\Pi = \frac{\eta v_k}{c_s},
\end{equation}
and we choose $\Pi = 0.05$, a typical value that applies to a wide variety of disc models \citep{Bai102}. 

The strength of self-gravity versus tidal shear is controlled by
\begin{equation}
\label{eq:Gtilde}
\widetilde{G} \equiv \frac{4\pi G \rho_{g,0}}{\Omega^2}.
\end{equation}
Selecting $\widetilde{G}=0.05$ is equivalent to a \citet{Toomre64} $Q$ of approximately 32, so the gas phase is gravitationally stable, supporting our exclusion of the gas density field in solving for the gravitational potential (equation~\ref{eq:poisson}).

In equation~\ref{eq:Gtilde}, $\rho_{g,0}$ is the gas midplane density. The gas density is initialized to have a Gaussian profile vertically with a scale height $H_g$, and a uniform distribution in the radial and azimuthal directions. The dust phase is initialized analogously except with a scale height of $H_d = 0.02 H_g$. We set the units of our scale-free model to that $\rho_{g,0} = H_g = \Omega = c_s = 1$. See Section 2.4 of \citet{Rucska21} for a discussion on how to convert these units to physical units. Using the minimum mass solar nebula model of \citet{Hayashi81} and placing our model at 3 AU, there is approx $1.5\ M_{\text{Ceres}}$ worth of dust mass in our full simulation domain.

The 3D simulation domains we study have equal lengths of $L_x=L_y=L_z=0.2\ H_g$, and we choose a grid resolution of $N_x=N_y=N_z=120$. \citet{Simon16} show that this resolution and \citet{Rucska21} show that this box size is sufficient to accurately capture the planetesimal formation process with our chosen set of physical parameters. This grid resolution matches that of \citet{Rucska21}.

We choose a dust resolution such that the total number of particles for each grain species is equal to the total number of grid points in the gas grid. The millions of dust particles are initially placed so that the overall dust density distribution is uniform in the $x$-$y$ plane and follows a Gaussian profile vertically. The precise initial positions of the particles are set via a random number generator. As in \citet{Rucska21}, we re-run multiple simulations that are otherwise identical except for the initial seed for the random number generator\footnote{The exact configuration of our runs, including this random seed, are available by request.}, which gives a different initial (and very small in amplitude) noise pattern to the dust density in each run. Once the streaming instability develops into the non-linear phase, the initial perturbations result in dramatic variations in the dust density. Thus, re-running simulations with different initial seeds probes the stochastic qualities of the non-linear SI and the variance in the outcomes in a way that a single simulation cannot.

In Table~\ref{tab:sims3}, the simulation labels \texttt{S0,...,S3} denote the simulations which use a single dust grain size (these are the same \texttt{L02(a-d)} simulations from \citealt{Rucska21}), and analogously the labels \texttt{M6-0,...,M6-4} represent five simulations that use multiple grain species simultaneously, with the only difference being the random seed that sets in the initial particle distribution. The \texttt{M12} and \texttt{M18} are simulations that sample the same size distribution as the \texttt{M6-0,...,M6-4} simulations but with a greater number of grain species/bins. Details on the grain sizes in each simulation are discussed in the proceeding section.

\subsection{Grain size distribution}
\label{sec:gsd}

We base our distribution of grain sizes on the results from \citet{Birnstiel11}, a widely used model of the collisional growth and fragmentation of dust grains in protoplanetary discs. Dynamics such as local turbulence, vertical settling, and radial drift affect the relative velocities between dust grains and can lead to grain growth or fragmentation via destructive collisions, depending on local conditions \citep[for a review, see][]{Birnstieletal16}. 

\citet{Birnstiel11} conclude that the dust grain population will equilibrate towards a size distribution with a shape that depends on properties of the disc (see their Fig. 6). Relevant properties include the gas surface density, midplane temperature, the \citet{ShakSun73} $\alpha$ turbulent viscosity parameter, and a fragmentation threshold velocity for the grains. The authors also provide an online tool for exploring different combinations of disc quantities. For the distribution shape we study, we choose $\Sigma_g=100$ g/cm$^2$, $T_{\text{mid}}=100$ K, roughly equivalent to a radial position of $\sim5$ AU for a disc with $\Sigma_g(r)= 1000\  (r/\text{AU})^{-3/2}$ $\text{g}/\text{cm}^2$ (e.g. minimum mass solar nebula model; \citealt{Weiden77-MMSN}) and $T_{\text{mid}} = 200\ (r/\text{AU})^{-3/7}$ K (e.g. \citealt{ChaGold97}). For other parameters we choose $\alpha=1\times 10^{-4}$, $v_{\text{frag}}=3$ m/s. These choices lead to a distribution that peaks around $\sim$4 cm at 5 AU.  For a lower gas surface density of $\Sigma_g=10$ g/cm$^2$, more indicative of typical radii observed by ALMA, the \citet{Birnstiel11} model predicts peak of size $\sim$3 mm. Thus, the physical grain size at the peak can change over an order of magnitude depending on disc properties, or with radial position in the disc.

The simulation impact of grain size is to set the characteristic time scale for the aerodynamic coupling between the dust and gas, $t_{\text{stop}}$ (equations~\ref{eq:gasmom} and~\ref{eq:dustmomprime3}). There are different forms for this stopping time depending on the regime of drag one considers, but for protoplanetary discs, almost all grains are in the Epstein \citep{Epstein24} drag regime \citep{Birnstieletal16}, where the size of the dust grains is smaller than the mean free path of the gas particles. The form of $t_{\text{stop}}$ in this regime is \citep{YG05, Bai101}
\begin{equation}
t_{\text{stop}} = \frac{\rho_s}{\rho_g c_s} s,
\end{equation}
where $\rho_s$ is the material density of the particles \citep[approximately $2.6$\,g\,cm$^{-3}$ for silicates;][]{Moore1973}, $\rho_g$ is the local gas density, $c_s$ is the local sound speed, which depends on the gas temperature, and $s$ is the size of the dust grains. Thus, for the same gas properties,  $t_{\text{stop}}$ scales linearly with grain size. In our models, as with other studies of the streaming instability, we model the drag coupling between the dust and gas with a dimensionless parameter $\tau_s=t_{\text{stop}} \Omega$,
\begin{equation}
\label{eq:taus}
\tau_s = \frac{\Omega\rho_s s}{\rho_g c_s}.
\end{equation}

In our disc model, the midplane gas density is $\rho_{g,0}=(1/\sqrt{2\pi})(\Sigma_g/H_g)$ and $H_g=c_s/\Omega$ \citep{Armitage}, so that $\tau_s = (\rho_s/\Sigma_g) s$. With our above choices for the disc properties in the grain size distribution (at 5 AU), the previously mentioned size peak of 4 cm grains translates to $\tau_s \sim 0.1$. In this study, we wish to directly compare our results to both our previous study \citep{Rucska21} and prior work which has focused on a single stopping time of $\tau_s=0.314$. Thus, we maintain the original shape of this particular \citet{Birnstiel11} distribution from our chosen disc parameters, but slightly shift the peak to $\tau_s=0.314$.  Within the model of \citet{Birnstiel11} this is equivalent to moving the fragmentation threshold from 3 km/s to $\sim$ 5 km/s or modifying the temperature profile.  Setting the peak at our previous single grain size value allows us to directly compare these two different representations of the dust environment.

As mentioned previously, the peak of the distribution from \citep{Birnstiel11} can vary strongly with the radial position within the disc. Broadly speaking, we expect our models to apply to $\sim$cm-sized grains in the inner 10 AU of the disc, and to $\sim$mm sized grains in the outer 10-100 AU regions of the disc, the latter being the kind of dust visible at ALMA wavelengths.

\subsubsection{Sampling the \citet{Birnstiel11} distribution}
\label{sec:sampbirn}

\begin{table}
 \caption{Dust mass in our sampled distributions.}
 \label{tab:sampbirn}
 \begin{tabular}{lccc}
  \hline
  Run names & 
  \multicolumn{3}{c}{\makecell{Dust mass in each $\tau_s$ bin, in units of the total dust mass.\\ Values are sorted by ascending $\tau_s$.}} \\
  
  \hline  
  
  \rule{0pt}{6.5ex}\noindent

  \makecell[l]{\texttt{M6-0,M6-1,M6-2,}\\\texttt{M6-3,M6-4}\\(\texttt{M6})} & 
  \multicolumn{3}{c}{\makecell{0.17582, 0.17594, 0.17535, \\ 0.16692, 0.15320, 0.15277}} \\

  \rule{0pt}{6.5ex}\noindent
  
  \texttt{M12} & 
  \multicolumn{3}{c}{\makecell{8.7975e-2, 8.7845e-2, 8.8100e-2, 8.7837e-2,\\ 8.7786e-2, 8.7563e-2, 8.3656e-2, 8.3265e-2,\\ 7.6858e-2, 7.6344e-2, 7.6577e-2, 7.6193e-2}} \\
  
  \rule{0pt}{9.5ex}\noindent
  
  \texttt{M18} & 
  \multicolumn{3}{c}{\makecell{5.8658e-2, 5.8687e-2, 5.8475e-2, 5.8736e-2 \\ 5.8746e-2, 5.8455e-2, 5.8627e-2, 5.8610e-2 \\ 5.8111e-2, 5.5892e-2, 5.6064e-2, 5.4965e-2 \\ 5.1262e-2, 5.1410e-2, 5.0530e-2, 5.1117e-2 \\ 5.1028e-2, 5.0625e-2}} \\
  
  \hline 
  \rule{0pt}{3ex}\noindent
 \end{tabular}
\end{table}

\begin{figure}
	\includegraphics[width=\columnwidth]{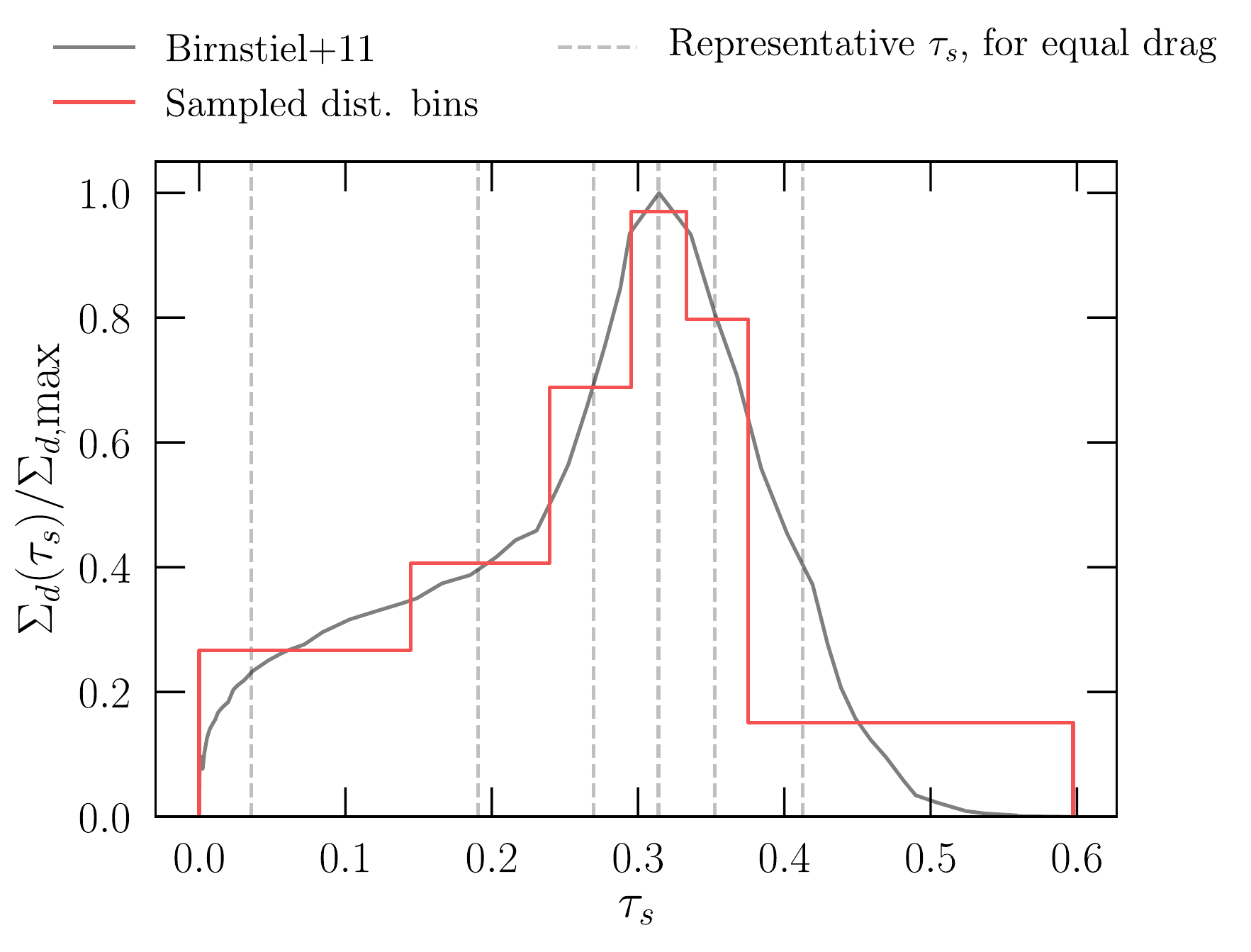}
    \caption[Sampled grain size distribution]{Grain size distribution sampled for this study. The grey curve represents the surface density distribution as a function of grain size according to a grain growth model in collision-fragmentation equilibrium \citep{Birnstiel11}. The red curve represents our sampling of the distribution with six bins, and the vertical dashed lines are the representative $\tau_s$ selected for each bin (See Section~\ref{sec:sampbirn}).}
    \label{fig:birndist}
\end{figure}

Figure~\ref{fig:birndist} shows the dust mass surface density distribution that we sample, as a function of $\tau_s$. The six different bins (red curve), used for the five \texttt{M6-0...M6-4} simulations, are chosen such that there is roughly equal dust mass in each bin while enforcing that one bin is centred on the peak at $\tau_s=0.314$\footnote{We could not simultaneously ensure that our sample has one bin with a representative $\tau_{\text{bin}}$ exactly at the peak $\tau_s=0.314$ and have the bins contain exactly equal mass. However all bin total masses are within 10\% of each other, as shown in Table~\ref{tab:sampbirn}.}. Since the streaming instability is driven by the aerodynamic coupling between the dust and gas, we chose the representative $\tau_s$ for each bin so that there is the same total drag force (proportional to $\Sigma_d/\tau_s$) in each bin as in the original distribution. This requires us to satisfy the equality
\begin{equation}
\label{eq:equaldrag}
\int_{\tau_{\text{bin,l}}}^{\tau_{\text{bin,r}}}\frac{\Sigma_d(\tau_s)}{{\tau_s}^{\prime}} d{\tau_s}^{\prime} = \frac{\Sigma_{\text{bin}}}{\tau_{\text{bin}}}\big( \tau_{\text{bin,r}} - \tau_{\text{bin,l}} \big),
\end{equation}
where $\tau_{\text{bin,(r,l)}}$ are the right and left $\tau_s$ values in each bin, $\Sigma_{\text{bin}}$ is the mean height of the distribution in that bin, and $\tau_{\text{bin}}$ is the representative size in that bin which we solve for. We see from Figure~~\ref{fig:birndist} that $\tau_{\text{bin}}$ in the bins for $\tau_s>0.1$ roughly tracks the half-mass point of the $\Sigma_d(\tau_s)$ curve, but in the bin for the smallest grains, $\tau_{\text{bin}}$ is closer to the leftmost, small-$\tau_s$ edge of the bin, because the drag force per unit mass scales as $1/\tau_s$.

Table~\ref{tab:sims3} lists the exact values of $\tau_{\text{bin}}$ (hereafter just referred to by $\tau_s$) modelled simultaneously by our simulations, with each species given a roughly equal amount of the total dust mass in the simulation domain. Note that these values of $\tau_s$ are not equally spaced, linearly or logarithmically, which is different from the distributions modelled by prior work \citep{Johansenetal07, Bai102, Schaffer18, YangZhu21} which also used equal-mass bins in their discretized distributions.

\subsubsection{Increasing the number of grain species}
\label{sec:meth-moregrains}

To accompany our main \texttt{M6-0...M6-4} simulations which use 6 bins, we also run two simulations with more species of grains/bins in order to test how our results are affected by the number of species present. We ran one with 12 bins and the other with 18 bins, which we denote \texttt{M12} and \texttt{M18} respectively. When creating these samples with additional bins, we decide to subdivide each of the original 6 bins into 2 and 3 bins, again keeping an equal mass in each bin. This maintains the original bin edges from the 6-bin sample and thus allows for a more straightforward comparison of the results between the different discrete distributions. Once the new (additional) bin edges are computed, the same procedure of equal drag from eqaution~\ref{eq:equaldrag} is used to select a representative $\tau_s$ for each bin.

\subsection{Planetesimal/clump identification}
\label{sec:clumping}

To quantify how the different sized grains participate in the formation of planetesimals in our simulations, we must first identify which grains are a part of bound planetesimals. For this study, we accomplish this with a dust density cut. The Hill radius denotes a region where the gravity of an object in a circumstellar disc dominates over the shear due to the velocity gradient of the background Keplerian rotation. Gravitational collapse is directly opposed by this rotational shear at large length scales and by diffusion of stirred up dust material at small scales \citep{Gerbig20,Gerbig23}. As described in \citet{Rucska21}, we can covert the Hill radius into a Hill density, above which a dust clump overcomes shear and is unstable to gravitational collapse. In the physical parameters of our model, this Hill density is given by,
\begin{equation}
\label{eq:rhoH}
\rho_H = 9 \frac{\Omega^2}{4\pi G}.
\end{equation}
With our choices of parameters, $\rho_H = 180$.

We identify all particles within cells with dust densities greater than $\rho_H$ as being a part of bound planetesimals, and all adjacent cells above this threshold are considered the same planetesimal. The triangular shaped cloud scheme that translates particle data to the gas grid smooths the dust density on the length scale of a single grid cell. As a result, some cells with relatively few particles have a dust density above $\rho_H$ because there are tens of thousands of particles in the neighbouring cells. We also average clump-related data over the multiple \texttt{M6-0...M6-4} simulations, removing some of the influence of the stochastic nature of the non-linear SI from our results concerning planetesimals. In this paper, we are not interested in the details of the clump mass distributions so we do not opt for a more sophisticated clump finding algorithm as in \citet{Rucska21}. 

% Results section 0
\section{Dust surface density at different grain sizes}
\label{sec:res_surfden}

\begin{figure*}
	\includegraphics[width=0.9\textwidth]{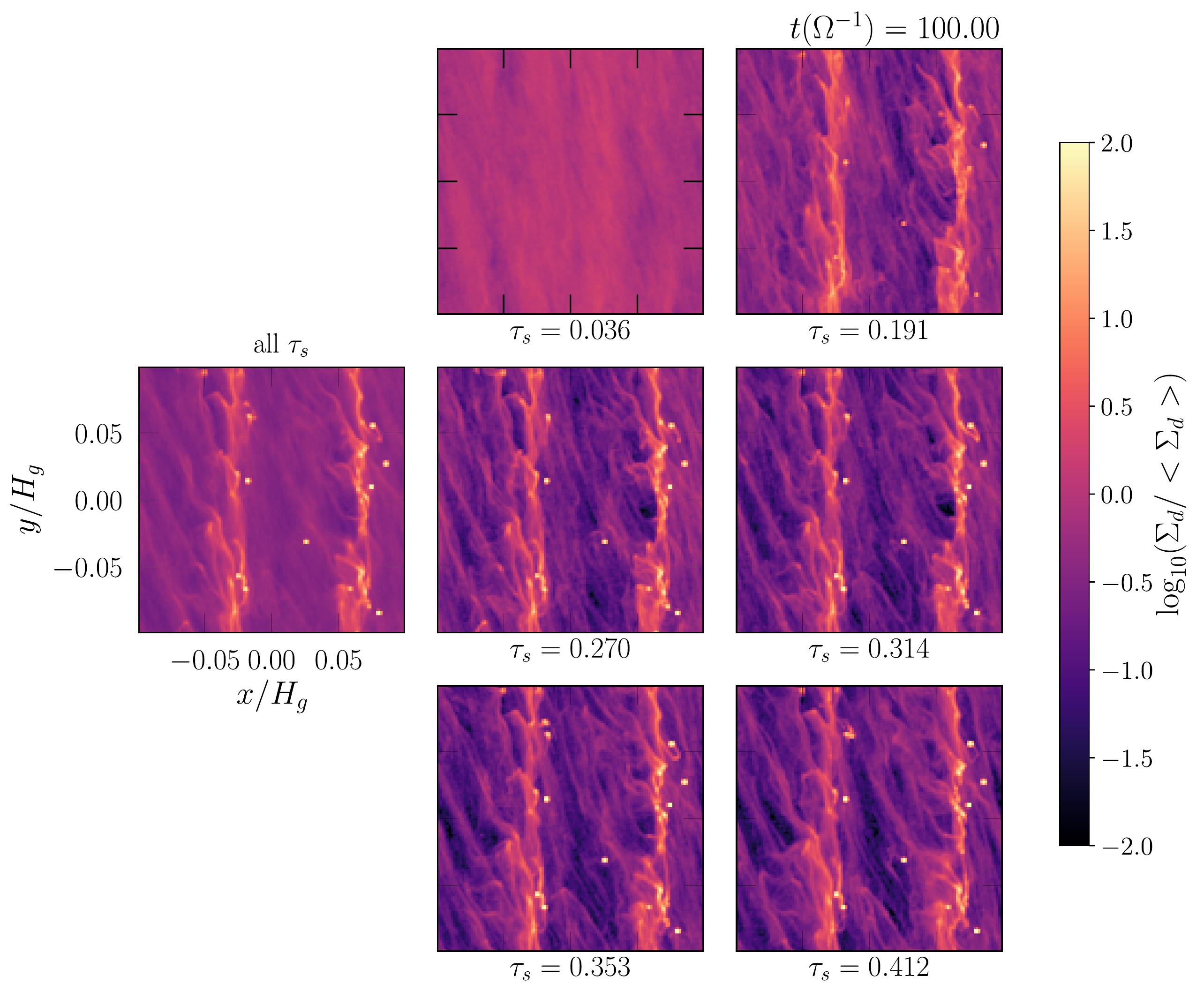}
    \caption[Dust surface density in the radial-azimuthal plane for each species of grain in the \texttt{M6-0} simulation]{Dust surface density in the $x$-$y$ (radial-azimuthal) plane for each species of grain in the \texttt{M6-0} simulation. The two right columns represent the surface density in the individual species, each identified by their grain size which here is represented by the dimensionless stopping time, $\tau_s$ (see equation~\ref{eq:taus} and surrounding discussion). The lone panel in the left column represents the total dust surface density in the simulation, with all grain species. The colour represents the logarithm of the dust surface density normalized by the mean dust surface density. The mean and normalization is computed in each panel individually. These data represents the simulation at time $t=100$ in units of the inverse orbital frequency, $\Omega^{-1}$.}
    \label{fig:surfden_spec}
\end{figure*}

\begin{figure*}
	\includegraphics[width=\textwidth]{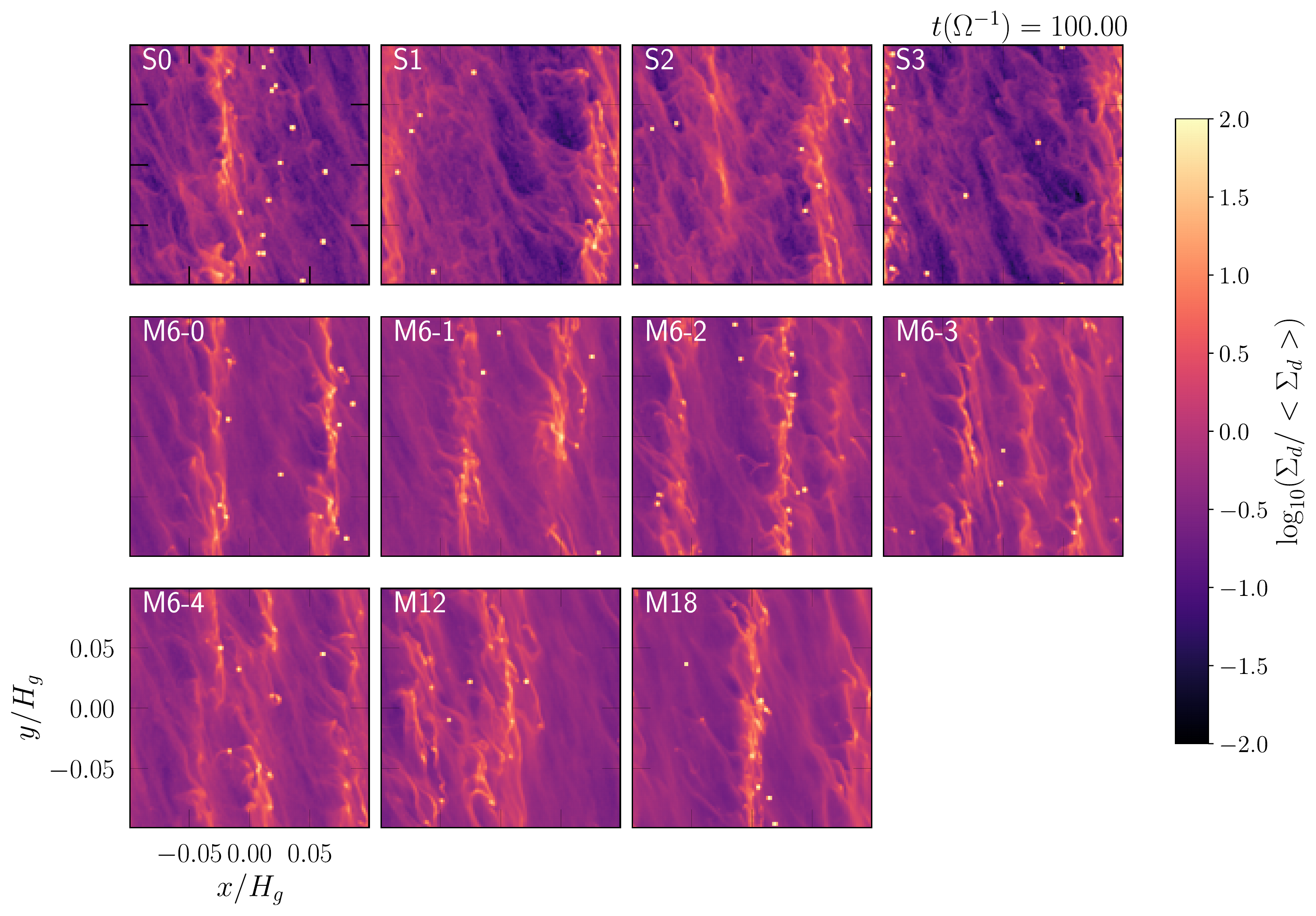}
    \caption[Dust surface density for all simulations]{Dust surface density at $t=100\ \Omega^{-1}$ in the $x$-$y$ (radial-azimuthal) plane for all simulations. The \texttt{S} simulations use a single grain size, and the \texttt{M6} simulations use multiple sizes simultaneously. See Table~\ref{tab:sims3} for a summary of the simulation parameters.}
    \label{fig:surfdenallsim}
\end{figure*}

\begin{figure}
	\includegraphics[width=\columnwidth]{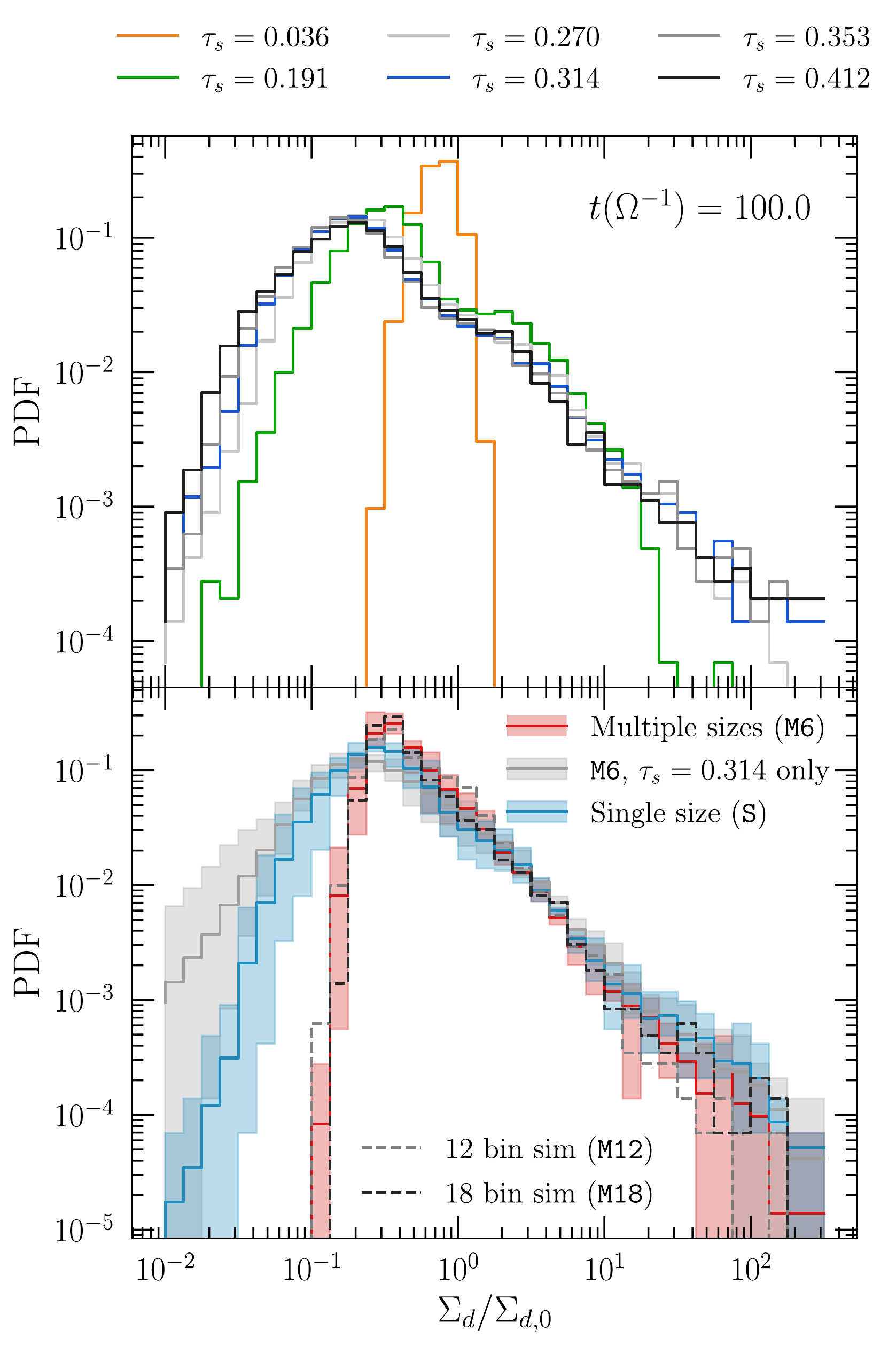}
    \caption[Probability distribution functions of the dust surface density]{Probability distribution functions (PDFs) of the dust surface density in our simulations at $t=100 \Omega^{-1}$. \textit{Top.} PDF for the \texttt{M6-0} simulation for each $\tau_s$ (grain size) bin (cf. Figure~\ref{fig:surfden_spec}). The normalization by the mean surface density is computed for each grain species individually. \textit{Bottom.} PDFs for all simulations (cf. Figure~\ref{fig:surfdenallsim}). The red (blue) shaded regions represent the maximum and minimum bounds among all \texttt{M6} (\texttt{S}) simulations, and solid line represents the mean PDFs. The grey shaded data and solid line represent the $\tau_s=0.314$ grains from the \texttt{M6} simulations only. The dashed curves are the PDFs for the \texttt{M12} and \texttt{M18} simulations.}
    \label{fig:sigdPDFs}
\end{figure}

In this section we examine the dust surface density in the multiple-grain simulations (\texttt{M6-0,...,M6-4}, hereafter referred to collectively as \texttt{M6}) and compare them with the surface density from the single-grain simulations (\texttt{S0,...,S3}, hereafter \texttt{S}). We inspect the surface density maps visually and then present a quantitative analysis of rudimentary observational consequences resulting in differences from these maps.

The dust surface density in the 6 different sizes or species of dust grains at $t=100 \Omega^{-1}$ in the \texttt{M6-0} simulation is shown in Figure~\ref{fig:surfden_spec}. We present all data from single snapshots at $t=100 \Omega^{-1}$ because at this stage planetesimal formation has begun in earnest, but the planetesimals have not yet disrupted the other features in the dust such as the filaments. As we discuss in detail in Section 4.1 of \citet{Rucska21}, the numerical cross-sections of the planetesimals in our simulations (and all similar simulations in the literature) are unphysically large, which causes the planetesimals to post-formation interact more strongly with the other dust particles than we would expect in nature. Thus, the true final state of the dust surface density post planetesimal formation is uncertain. We pick $t=100 \Omega^{-1}$ as a compromise to capture the coexistence of the planetesimals and the filaments, which we expect to be typical of the saturated stage of the non-linear streaming instability.

We notice immediately in Figure~\ref{fig:surfden_spec} that the smallest sized dust grains (lowest $\tau_s$) do not readily collect into filaments or planetesimals at all, even when the larger grains are producing dense features simultaneously. All grains with $\tau_s > 0.1$ participate in the structure of the filaments, while the distribution of the $\tau_s=0.0355$ grains is smooth with relatively little spatial variation. Secondly, with a more careful visual inspection of the $\tau_s=0.191$ surface density map, one can see that the brightest, $\sim$2-3 cell wide objects in the largest grains--which represent the planetesimals--are less bright than in the $\tau_s > 0.2$ grains, suggesting the $\tau_s=0.191$ grains do not incorporate into planetesimals as readily (for more quantitative results concerning clumping, see Section~\ref{sec:res_clumps}). In the full dust surface density, which includes all grains (``all $\tau_s$'' panel), we see altogether the filaments, planetesimals, and the smooth, dispersed quality of the smallest grains which is most apparent in the space between filaments. Similar visual features can be seen in other studies of the non-linear SI with multiple grain sizes (cf. \citealt{YangZhu21} Figures 4 and 6, \citealt{Bai102} Figure 2, \citealt{Johansenetal07} Figure 2).

We can make similar observations when comparing the surface density maps of the \texttt{M6} and \texttt{S} simulations, shown in Figure~\ref{fig:surfdenallsim}. The \texttt{S} cases use one grain size of $\tau_s=0.314$ and thus more closely resemble the $\tau_s=0.270$ to $0.412$ grains from the \texttt{M6} simulations, in that the dust mass at these sizes is predominantly concentrated into planetesimals and filaments which are separated by relatively empty regions with surface densities $\lesssim$10\% of the mean surface density. The smaller grains in the \texttt{M6} simulations fill these empty regions.

More quantitative confirmation of these observations can be seen in the probability distribution functions (PDFs) of the dust surface density, in Figure~\ref{fig:sigdPDFs}. The top panel shows the PDFs for the individual grains from the \texttt{M6-0} simulation at $t=100 \Omega^{-1}$. Here, we quantify what is observable in Figure~\ref{fig:surfden_spec}: the distribution of surface density in the $\tau_s=0.036$ grains is narrow, peaking around the mean, $\langle\Sigma_{d}\rangle$. The $\tau_s=0.191$ distribution is wider by an order of magnitude in each direction, which is a sign that these grains are participating in filaments. Yet only the grains with $\tau_s>0.2$ extend out to surface densities greater than $100\, \langle\Sigma_{d}\rangle$ -- a (rough) proxy for planetesimals. The PDFs of the particle volume density from \citet{YangZhu21} Figure 8 show similar segregation by grain size.

The bottom panel of Figure~\ref{fig:sigdPDFs} highlights how this affects the overall surface density in the \texttt{M6} simulations. The PDFs extend only as low as $0.1\,\langle\Sigma_{d}\rangle$, while the distributions from the single size $\tau_s=0.314$ simulations extend out to $0.01\,\langle\Sigma_{d}\rangle$. Interestingly, when looking at the $\tau_s=0.314$ grains from the \texttt{M6} simulations on their own, these PDFs show there are more low surface density areas in these grains than there are in the \texttt{S} simulations at this size. This suggests that the presence of different-sized dust grains in the \texttt{M6} results in more empty or lower surface density regions (in the $x$-$y$ plane) than if the $\tau_s=0.314$ grains were left to evolve on their own. A possible physical interpretation for this observation is that, since both the \texttt{M6} and \texttt{S} models have the same total dust mass in their simulation domains, there is less dust mass in SI-active grains ($\tau_s \gtrsim 0.3$) in the \texttt{M6} domains than the \texttt{S} domains. All the dust in the \texttt{S} runs has $\tau_s=0.314$, which generates turbulence via the SI which can diffuse the dust mass \citep{Gerbig20, Gerbig23}. This results in low surface density regions (in the $x$-$y$ plane) being less common in the \texttt{S} model ($\tau_s=0.314$) dust when compared to the \texttt{M6} model $\tau_s=0.314$ dust. (c.f. the surface density maps for \texttt{S0}-\texttt{S4} in Fig.~\ref{fig:surfdenallsim} and the $\tau_s=0.314$ grains for the \texttt{M6-0} run in Fig.~\ref{fig:surfden_spec}.) We discuss these ideas further in Section~\ref{sec:vertdyn}, where we investigate the vertical (out-of-midplane) velocity dispersion in the dust.

The PDFs for the \texttt{M12} and \texttt{M18} simulations--which have more grain size bins than the \texttt{M6} simulations--are also shown in the bottom panel of Figure~\ref{fig:sigdPDFs}. These PDFs follow the \texttt{M6} data closely, suggesting that six grain species was sufficient to capture the main effects on the dust surface density.

The differences in the \texttt{S} (blue) and \texttt{M6} (red) PDFs have interesting observational consequences. There many more regions with low surface density in the \texttt{S} simulations and (on average) more regions at higher surface densities. Depending on the opacity of the dust, this could lead to a lower estimate of the total dust mass from observations due to optical depth effects. We explore this idea in the next section.

\subsection{Observational consequences}
\label{sec:obscon}

\begin{figure}
	\includegraphics[width=\columnwidth]{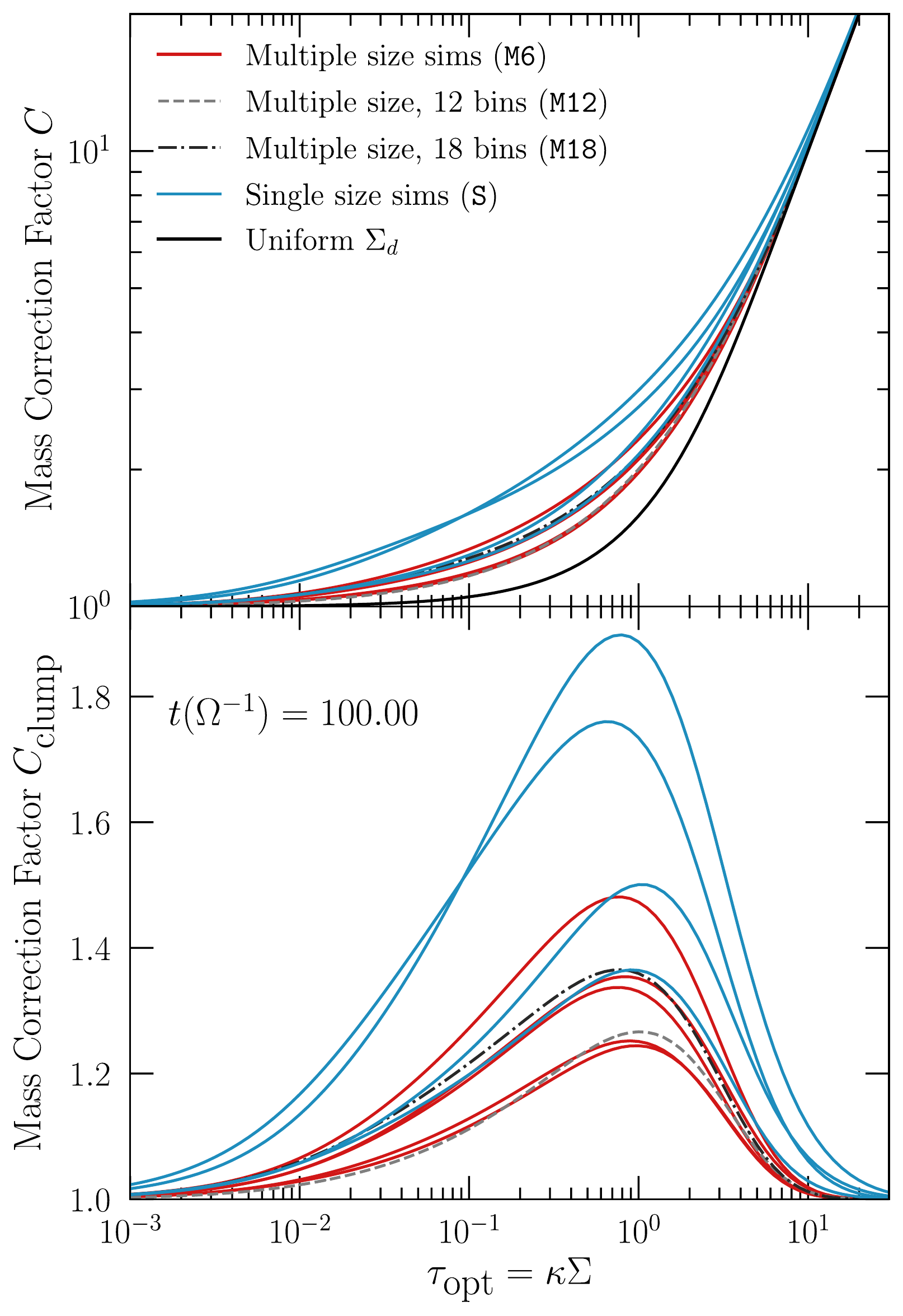}
    \caption[Observational dust mass correction factors]{Observational dust mass correction factors.  \textit{Top.} $C$: The ratio of the true surface density ($\Sigma_{d}$) to an estimate that assumes the disc to be optically thin with no unresolved structures within the beam, as a function of the mean optical depth. \textit{Bottom.} $C_{\rm clump}$:  The correction factor due to unresolved clumping from the streaming instability alone, with uniform optical depth effects removed.  These curves are equivalent to dividing the coloured and dashed curves in the top panel by the solid black curve in the top panel. \textit{Both.} Each blue or red curve represents an individual simulation. Details on how these factors are defined are in Section~\ref{sec:obscon}.}
    \label{fig:Cfacs}
\end{figure}

\begin{figure}
	\includegraphics[width=\columnwidth]{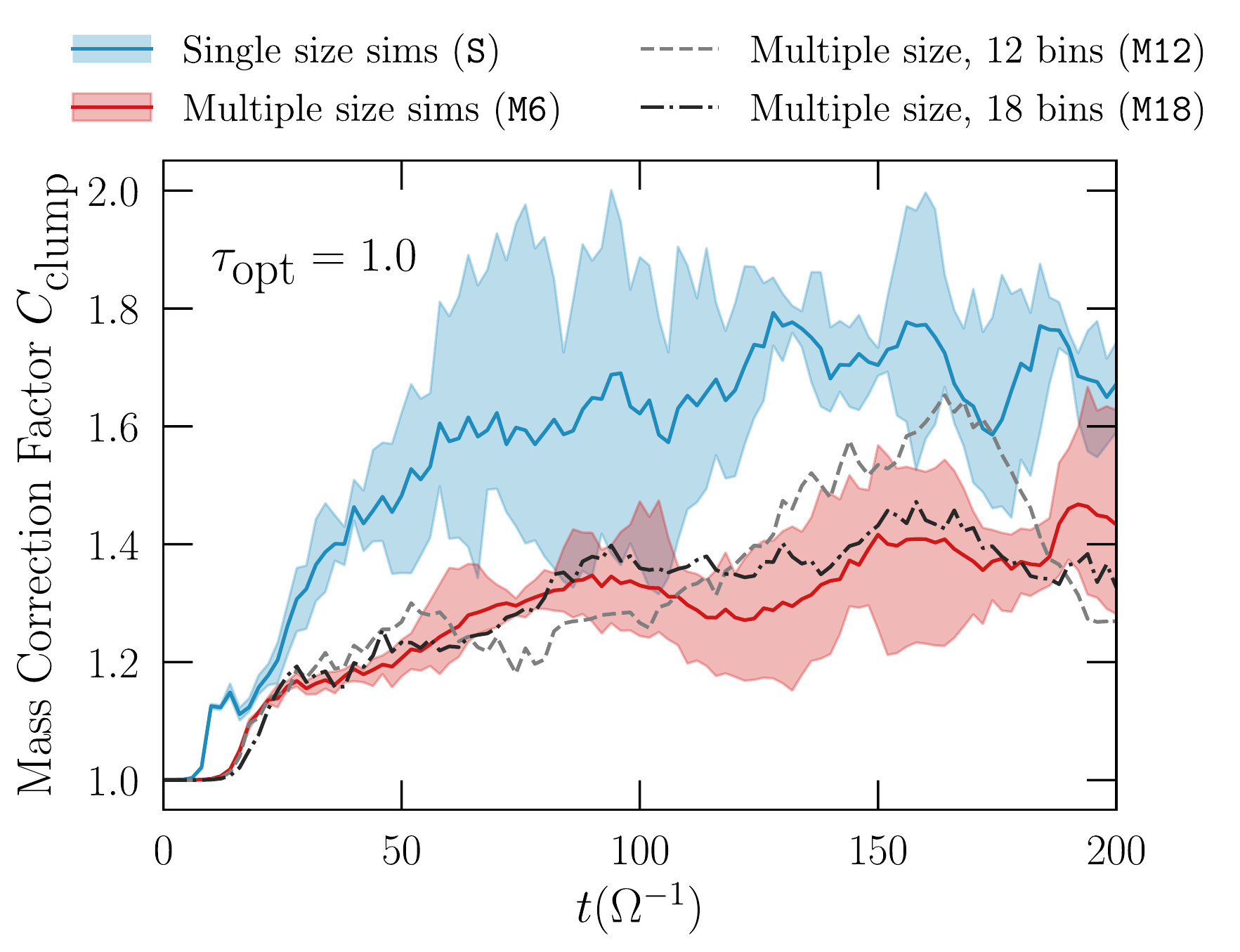}
	\caption[The dust mass correction factor over time]{The dust mass correction factor $C_{\text{clump}}$ over time in all simulations, at an optical depth of $\tau_{\textrm{opt}}=1.0$. The shaded regions are bounded by the maximum and minimum values across the sample of multiple simulations, and the solid curves represent the means of that sample.}
    \label{fig:Covertime}
\end{figure}

Observations of some bright rings in protoplanetary discs have come to the interesting conclusion that the thermal emission from the dust in these rings is likely not optically thick \citep{Dull18-DSVI, Huang18-DSII, Cazzoletti18, Macias19, Mauco21}. Other studies have shown that, with a parameterized model of planetesimal formation via the streaming instability, this can be explained by pebble-sized dust in rings being converted into planetesimals, which do not contribute to mm wavelength emission \citep{Stammler19, Mauco21}. Taking this idea a step further, \citet{Scardoni21} used the dust surface density profiles from 2D simulations of the SI and explored how the dust clumping would affect observations. They use a complex model for the dust opacity \citep{Birn18-DSV} and find general agreement between their calculations of observed properties of discs such as the fraction of the emission that is optically thick and the spectral index. Unsurprisingly, they conclude that planetesimal formation can reduce the optical depth of emission at mm wavelengths.

In this section, we construct two mass correction factors which quantify the observational implications of the varying degrees of dust clumping seen in our simulations\footnote{Note, the dust features created by the SI occur on length scales several orders of magnitude below 1 AU, and are hence unresolvable by any contemporary observational facility.}. We explore how these mass correction factors vary with optical depth ($\tau_{\text{opt}}=\kappa \Sigma_{d}$), and how they evolve over time.  Note that to leading order effects like disc inclination can be wrapped into a different effective $\kappa$.  We forgo a detailed mock observational treatment and complicated calculations of the dust opacity. Instead, we examine how the mean emission is modified due to both unresolved structure from the SI and differences between single grain size and multiple grain size models.

The intensity of emission ($I$) from a source of radiation (source function $S$, and other   dust physical properties assumed constant), can be expressed as,
\begin{equation}
    \label{eq:rt-I}
    I(\tau_{\text{opt}}) = S \big(1 - e^{-\tau_{\text{opt}}} \big),
\end{equation}
where $\tau_{\text{opt}}$ is the optical depth, and the wavelength dependence of all quantities has been ignored. We consider a simple prescription for the optical depth, $\tau_{\text{opt}}= \kappa \Sigma_d$, where $\kappa$ is the dust opacity and $\Sigma_d$ is the dust surface density. For optically thin emission (low opacity, and/or low surface density), $\tau_{\text{opt}} \ll 1$, and then $I \approx S\kappa \Sigma_d$, linear in the surface density. If $S$ and $\kappa$ are known, we can estimate  $\Sigma_{\text{est}} = I/S\kappa$. However, assuming the emission is optically thin systematically underestimates the surface density at points where $\tau_{\rm opt}$ is not small, as,
\begin{equation}
\label{eq:rt-sdest}
\Sigma_{\text{est}} = \frac{\big(1 - e^{-\kappa \Sigma_d} \big)}{\kappa}.
\end{equation}
This expression is useful as it does not require knowledge of the source function to assess the potential for systematic errors.

For an assumed value of $\tau_{\text{opt}}$, we take the surface density map from our simulations, $\Sigma_d(x,y)$, and compute $\Sigma_{\text{est}}(x,y)$ via equation~\ref{eq:rt-sdest}. For simplicity we assume a single opacity, $\kappa$, for all dust, independent of $\tau_s$. We take the spatial average of $\Sigma_{\text{est}}(x,y)$, using the fact that all features in our simulations would be unresolved within an observational beam.  Henceforth, we will refer to this average using, $\Sigma_{\text{est}}=\langle\Sigma_{\text{est}}(x,y)\rangle$.

We construct two dust mass correction factors from these average estimate surface densities. First, using a ratio of the true averaged dust surface density of the  simulation $\Sigma_{\text{actual}}$  (i.e. $\Sigma_{\text{actual}}=\langle\Sigma_d(x,y)\rangle$),
\begin{equation}
\label{eq:C1}
C = \frac{\Sigma_{\text{actual}}}{{\Sigma_{\text{est}}}}.    
\end{equation}
$C$ is an overall correction factor for optically thick, unresolved clumping.

To examining clumping alone, we can compute a new estimate, $\Sigma_{\text{est, uniform}}$,  assuming a \textit{uniform} dust density distribution, $\Sigma(x,y)=\langle\Sigma(x,y)\rangle$, and compare that to  $\Sigma_{\text{est}}$ from our strongly clumped simulations.  The ratio,
\begin{equation}
\label{eq:C2}
C_{\text{clump}} = \frac{\Sigma_{\text{est,uniform}}}{\Sigma_{\text{est}}}
= \frac{C}{C_{\text{uniform}}} ,   
\end{equation}
measures the correction associated with of dust clumping  only (here due to the streaming instability).
$C_{\text{uniform}}={\Sigma_{\text{actual}}}/{\Sigma_{\text{est,uniform}}}$ is the correction associated assuming low optical depths.
If one is confident about the optical depth of an observed source, $C_{\text{clump}} $ represents the factor the inferred dust surface density should be multiplied by if the SI is believed to have caused significant unresolved clumping in that region of the disc.

The top panel of Figure~\ref{fig:Cfacs} shows $C$ as a function of $\tau_{\text{opt}}$. At low $\tau_{\text{opt}}$, $C\sim 1$ for all simulations, since in this regime the optically thin assumption is valid by definition. At intermediate $\tau_{\text{opt}}$, expected variation between simulations \citep{Rucska21} leads to a spread in $C$. At high $\tau_{\text{opt}}$, all simulations converge to $C \sim \tau_{\text{opt}}$ since the exponential term in equation~\ref{eq:rt-sdest} vanishes, removing any dependence in $C$ on the surface density distribution in the simulations and hence any influence of clumping. It is at intermediate values of $\tau_{\text{opt}}$ where differences between the sets of simulations are apparent.

To understand the impact of clumping, we look at the clumping correction factor, $C_{\text{clump}}$, in the bottom panel of Figure~\ref{fig:Cfacs}. This factor is the ratio of $C$ for each simulation (coloured and dashed lines, top panel) to  $C_{\text{uniform}}$, calculated for a spatially uniform dust surface density distribution at $\,\langle\Sigma_{d}\rangle$ (black line, top panel\footnote{This curve is simply a plot of $\tau_{\text{opt}}/(1- \exp(-\tau_{\text{opt}}))$.}). Hence, what $C_{\text{clump}}$ highlights is the influence of the different amount of clumping in the dust surface density maps/distributions between the two sets of simulations (Figure~\ref{fig:surfdenallsim} and~\ref{fig:sigdPDFs}). If the optical depth is well constrained (e.g. from grain properties), it is the factor one would multiply a dust surface density inferred from observations by in order to account for the (unresolved) dust clumping from our simulations.

We note that $C_{\text{clump}}$ peaks near $\tau_{\text{opt}}=1$ for all simulations, with peak values generally higher for the single size simulations \texttt{S} than the \texttt{M6} simulations with multiple sizes. As we discussed earlier in Section~\ref{sec:res_surfden}, when compared the simulations with multiple grains, the dust in the \texttt{S} sims is more heavily concentrated into denser structures. As a consequence, in our simple model, the dust emission from the \texttt{S} models is overall less bright than the \texttt{M6} models, as there is relatively less dust mass in the inter-filament space, and there is more dust in the filaments and planetesimals, where the emission is saturated at intermediate optical depths. Thus more of the dust mass is ``hidden'' in the \texttt{S} simulations.

The peak values of $C_{\text{clump}}$ are between $\sim1.2-1.5$ for the \texttt{M6} and \texttt{M12} and \texttt{M18} simulations, which we believe to be more representative of protoplanetary disc grain size distributions in nature rather than a single size. Thus, for dust grains described by a \citet{Birnstiel11} grain size distributions with stopping times peaked at $\tau_s=0.314$, observational estimates of the dust mass from protoplanetary discs could be too low by a factor of $20-30\%$ in regions of the disc where the streaming instability is active. If the grain size distribution were instead much more strongly peaked at a single size--i.e., closer to the \texttt{S} models than \texttt{M6}--then the clumping mass correction factor could be as high as factor of two.

The values of $C_{\text{clump}}$ for a mean optical depth,$\tau_{\text{opt}}=1.0$, over time, plotted in Figure~\ref{fig:Covertime}, are fairly stable over the course of our simulations. Once planetesimal formation begins, the main features of the dust surface density which influence $C_{\text{clump}}$ persist over dozens of dynamical timescales. Note that the single snapshot values of $C_{\text{clump}}$ presented in Figure~\ref{fig:Cfacs} are from $t=100 \, \Omega^{-1}$, and in Figure~\ref{fig:Covertime} this is one of the rare times where there is slight overlap between the single size and multiple size sims. At most other times the curves in Figure~\ref{fig:Cfacs} do not overlap at all. Similar to the single snapshot data, the curves for the \texttt{M12} (12 bins) and \texttt{M18} (18 bins) simulations in Figure~\ref{fig:Covertime} are consistent with the \texttt{M6} simulations, suggesting incorporating more grain species does not influence our results.

Note, we do not plot $C$ over time as it has the same shape of $C_{\text{clump}}$, since the difference in normalization between $C$ and $C_{\text{clump}}$ (at a specific optical depth) are just different constant factors. $C$ uses $\,\langle\Sigma_{d}\rangle$ as a reference point, and, for the purposes of Figure~\ref{fig:Covertime}, $C_{\text{clump}}$ uses $\langle\Sigma_{d,\text{unif}}\rangle(\tau_{\text{opt}}=1.0)$.

% Results section 1
\section{Planetesimal composition: grain size}
\label{sec:res_clumps}

\begin{figure}
	\includegraphics[width=\columnwidth]{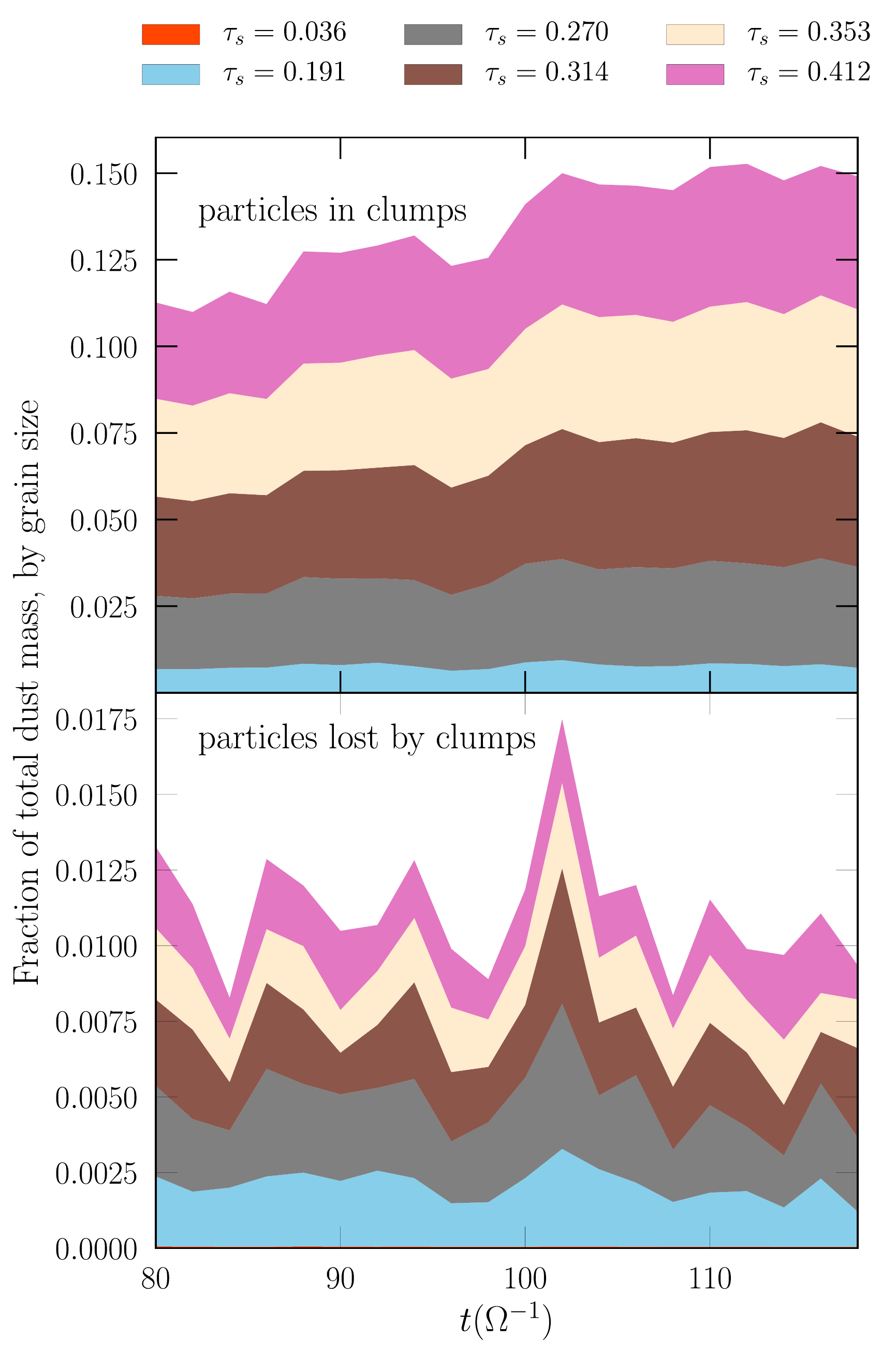}
	\caption[Fraction of total dust mass for particles in bound clumps or lost by clumps]{Fraction of total dust mass for particles in bound clumps or lost by clumps, for each $\tau_s$ (grain size) bin. These data represent an average over the whole group of \texttt{M6} simulations. The coloured bands represent the fractional mass for each $\tau_s$. The data for each grain size are vertically stacked so that the total mass in clumps (or lost by clumps) for all dust grains is tracked by the top of the pink shaded region. The data for $\tau_s=0.036$ are too small to be seen on this scale; see Table~\ref{tab:massclumps} for time-averaged values of these data for all $\tau_s$, and for the data from the \texttt{M12} and \texttt{M18} simulations.}
    \label{fig:fracmassot}
\end{figure}

\begin{table}
    \caption[Time averages of the total dust mass in bound clumps and lost by clumps]{Time averages ($t=80-120\Omega^{ -1}$) of the total dust mass in bound clumps and lost by clumps, split by $\tau_s$ (grain size) (cf. Figure~\ref{fig:fracmassot}).}
    \label{tab:massclumps}
	\begin{subtable}{.5\textwidth}
	\centering
	\subcaption{Dust mass in bound clumps (as fraction of total dust mass).}
	\label{tab:massclumpsa}
	\begin{tabular}{lccc} % four columns, alignment for each
		\hline
		$\tau_s$ & 6 bin runs & 12 bin & 18 bin \\
		\hline
        \rule{0pt}{3ex}\noindent
		0.036 & $4.90\times 10^{-5}$ & $4.38\times 10^{-4}$ &
		    $5.84\times 10^{-4}$\\
		0.191 & $7.73\times 10^{-3}$ & $7.47\times 10^{-3}$ &
		    $1.08\times 10^{-2}$\\
		0.270 & $2.60\times 10^{-2}$ & $1.81\times 10^{-2}$ &
		    $3.20\times 10^{-2}$\\
		0.314 & $3.38\times 10^{-2}$ & $2.29\times 10^{-2}$ &
		    $4.23\times 10^{-2}$\\
		0.353 & $3.30\times 10^{-2}$ & $2.24\times 10^{-2}$ &
		    $4.33\times 10^{-2}$\\
		0.412 & $3.43\times 10^{-2}$ & $2.06\times 10^{-2}$ &
		    $4.46\times 10^{-2}$\\
		\hline
	\end{tabular}
	\end{subtable}
	\rule{0pt}{14ex}\noindent
    \begin{subtable}{.5\textwidth}
	\centering
	\subcaption{Dust mass lost by clumps (as fraction of total dust mass).}
	\label{tab:massclumpsb}
	\begin{tabular}{lccc} % four columns, alignment for each
		\hline
		$\tau_s$ & 6 bin runs & 12 bin & 18 bin \\
		\hline
        \rule{0pt}{3ex}\noindent
		0.036 & $4.71\times 10^{-5}$ & $2.74\times 10^{-4}$ &
		    $3.52\times 10^{-4}$\\
		0.191 & $2.04\times 10^{-3}$ & $1.99\times 10^{-3}$ &
		    $2.70\times 10^{-3}$\\
		0.270 & $2.78\times 10^{-3}$ & $2.51\times 10^{-3}$ &
		    $3.83\times 10^{-3}$\\
		0.314 & $2.43\times 10^{-3}$ & $2.21\times 10^{-3}$ &
		    $3.38\times 10^{-3}$\\
		0.353 & $1.95\times 10^{-3}$ & $1.75\times 10^{-3}$ &
		    $2.73\times 10^{-3}$\\
		0.412 & $1.93\times 10^{-3}$ & $1.72\times 10^{-3}$ &
		    $2.55\times 10^{-3}$\\
		\hline
	\end{tabular}
	\end{subtable}
\end{table}

In this section we explore the composition of these clumps in terms of the various dust species within them, as well as the composition of the dust mass that lost from each clump from simulation snapshot to snapshot.  The bright cells in the surface density maps in Figure~\ref{fig:surfdenallsim} indicate that all simulations from our study produce dense, gravitationally bound clumps. As described in Section~\ref{sec:clumping}, we identify bound clumps (i.e. planetesimals) as regions where the 3D dust volume density ($\rho_d$) exceeds the Hill density ($\rho_H$) threshold above which it is unstable to gravitational collapse. 

The fraction of mass in clumps for each grain size is shown in Figure~\ref{fig:fracmassot}. The different coloured bands represent the mass in each grain size bin. Table~\ref{tab:massclumps} shows the time averages for these data over the range of time across the full $x$-axis in Figure~\ref{fig:fracmassot}. As seen in the top panel and in Table~\ref{tab:massclumpsa}, the majority of the mass in clumps ($>90\%$) is in the grains with $\tau_s>0.2$. A small fraction of the clumps are composed of $\tau_s=0.191$ grains and there is effectively no clump mass associated with the $\tau_s=0.035$ grains. These results corroborate earlier observations from Figure~\ref{fig:surfden_spec} regarding the decreased prominence or total lack of visible planetesimals in the surface density maps for these grain sizes.

Some particles that are within in a clump in one snapshot are not within that same clump\footnote{Planetesimals in concurrent simulation snapshots which share over $50\%$ of the same unique particles (determined by particle ID numbers) are determined to be the same clump.} in the consecutive snapshot. These particles may be loosely bound at the edge of the gravitational influence of the planetesimal (i.e. near the Hill radius) or simply passing through the high-density grid cells that are identified as planetesimals. We will explore these ideas with velocity and vertical position data in Section~\ref{sec:vel-zpos}. For the purposes of this analysis, we identify these transient clump particles as ``lost'', and plot the composition of this lost dust mass in the bottom panel of Figure~\ref{fig:fracmassot} and provide the time averages of these data in Table~\ref{tab:massclumpsb}. The lost dust mass is nearly evenly distributed among the grains at $\tau_s > 0.1$, with the highest proportion involving the $\tau_s=0.270$ grains. Note that on average, approximately 10\% of all the mass in clumps is consistently lost between snapshots.

\begin{table}
	\centering
	\caption[Residence times for the \texttt{M6} simulations.]{Residence time (equation~\ref{eq:restime}) for the different dust grains in the \texttt{M6} simulations.}
	\label{tab:restime}
	\begin{tabular}{lccc} % four columns, alignment for each
		\hline
		$\tau_s$ & Mass & Mass lost & Residence \\
		\ & in clumps & from clumps &  time ($\Omega^{-1}$)\\
		\hline
        \rule{0pt}{3ex}\noindent
		0.036 & $4.90\times 10^{-5}$ & $4.71\times 10^{-5}$ & 2.08\\
		0.191 & $7.73\times 10^{-3}$ & $2.04\times 10^{-3}$ & 7.60\\
		0.270 & $2.60\times 10^{-2}$ & $2.78\times 10^{-3}$ & 18.7\\
		0.314 & $3.38\times 10^{-2}$ & $2.43\times 10^{-3}$ & 27.8\\
		0.353 & $3.30\times 10^{-2}$ & $1.95\times 10^{-3}$ & 33.9\\
		0.412 & $3.43\times 10^{-2}$ & $1.93\times 10^{-3}$ & 35.6\\
		\hline
	\end{tabular}
\end{table}

We can combine the results from the two panels of Figure~\ref{fig:fracmassot} into a single idea known as the residence time--a quantity that estimates how long the dust mass of a particular grain species will remain in clumps given how quickly that mass is lost. This is represented simply by,
\begin{equation}
\label{eq:restime}
\text{Residence time} = \Delta t_{\text{snap}}
  \Bigg( \frac{\text{Mass in clumps}}{\text{Mass lost btwn. snapshots}} \Bigg),
\end{equation}
where $\Delta t_{\text{snap}}$ is the amount of time between data outputs and in this study is equal to $2.0\ \Omega^{-1}$. The residence time is hence equivalent to diving the data in Table~\ref{tab:massclumpsa} by the data in Table~\ref{tab:massclumpsb} and multiplying by $\Delta t_{\text{snap}}$.

We present calculations of the residence time in Table~\ref{tab:restime}. This table confirms our prior conclusions when considering both panels of Figure~\ref{fig:fracmassot} together: the largest grains are the most bound, longest-lived components of the planetesimals. All grains with $\tau_s > 0.2$ have residence times above $18\ \Omega^{-1}$, and this quantity increases monotonically with $\tau_s$. The smallest grains at $\tau_s =0.036$ have residence times comparable to $\Delta t_{\text{snap}}$, suggesting they form only a transient component of the clump mass\footnote{A more sophisticated clump-finding approach may definitely determine these small grains to be kinematically unbound. However, our simpler (and less expensive) analysis reaches the same conclusion to the degree of precision suitable for our study.}. 
The approximate gravitational free-fall time for a bound cloud of dust grains to collapse to asteroid densities (i.e. a planetesimal) is of order $\Omega^{-1}$. Thus, so long as the residence time of a dust species is $\gg\ \Omega^{-1}$, that dust should be easily incorporated into planetesimals.

Interestingly, the $\tau_s=0.191$ grains have an intermediate residence time of $\sim8\ \Omega^{-1}$. We can also observe from the dust surface density maps for each grain species (Fig.~\ref{fig:surfden_spec}) and the PDF of those surface densities (top panel Fig.~\ref{fig:sigdPDFs}) that the $\tau_s=0.191$ grains exhibit behavior that is not like the smallest grains or the larger grains. The smallest grains do not participate in any kind of dust clumping, and the larger grains readily form gravitationally unstable planetesimals. Our results suggest the $\tau_s=0.191$ grain behavior is in-between these two regimes.

\begin{figure}
	\includegraphics[width=\columnwidth]{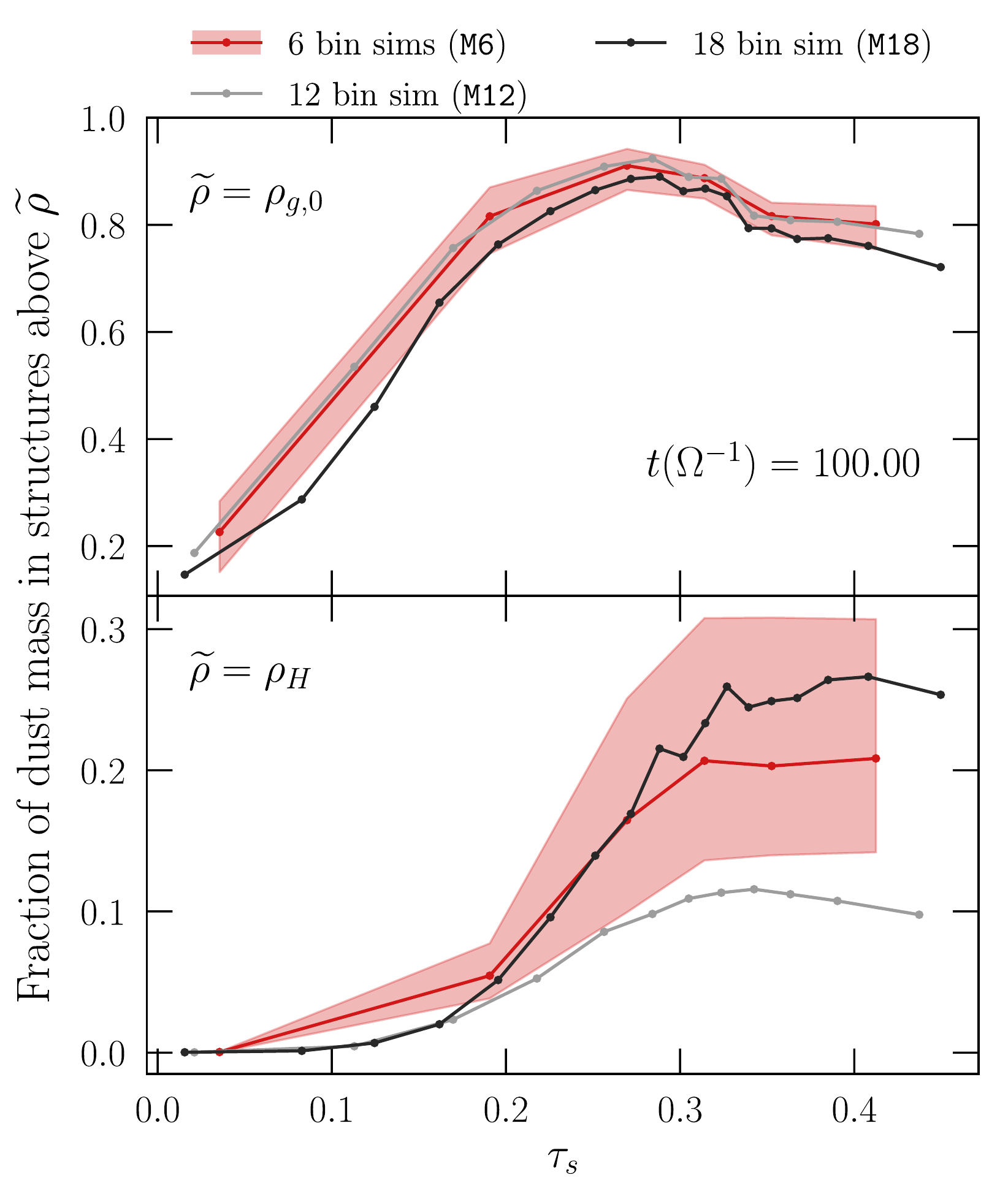}
	\caption[Total dust mass above certain density thresholds as a function of grain size]{Total dust mass above certain density thresholds ($\widetilde{\rho}$) as a function of grain size ($\tau_s$), normalized by the total dust mass in each grain size bin. In the top panel the threshold is the initial midplane gas density $\rho_{g,0}$ and the bottom panel the threshold is the Hill density (equation~\ref{eq:rhoH}), the threshold above which dust forms gravitationally bound planetesimals. The shaded regions represent the bound for the maximum and minimum across the five \texttt{M6} simulations.}
    \label{fig:smplot}
\end{figure}

We can see evidence of this in-between behavior for the $\tau_s=0.191$ grains in Figure~\ref{fig:smplot}, which shows the amount of dust mass above a certain density threshold at each grain size at $t=100\ \Omega^{-1}$. In the bottom panel, the threshold is $\rho_H$, and hence these data are equivalent to (a single time/vertical slice of) the data from the top panel of Figure~\ref{fig:fracmassot}. We see similar conclusions as before: the $\tau_s > 0.2$ grains dominate the clump mass budget, the $\tau_s=0.036$ grains are not a part of the clumps at all, and the $\tau_s=0.191$ make up a small fraction of the mass at clump densities. These results agree with widely-cited studies from the literature which, for the dust-to-gas surface density ratio in our model, identify $\tau_s \sim 0.1$ to $0.3$ as the most unstable stopping time/grain size for strong clumping and planetesimal formation via the SI \citep{Carrera15, Yang17, Li21}.

In the top panel of Figure~\ref{fig:smplot}, the threshold is $\rho_{g,0}$, the mid-plane gas density. In our simulations and those like it from the literature, the gas density displays little variation, even when the streaming instability develops strong dust clumps and filaments \citep{Li18}. So the $\rho_{g,0}$ threshold effectively marks the boundary where the dust density dominates the total local mass density ($\rho = \rho_d + \rho_g$), an important regime for the streaming instability \citep{YG05}. We see that the $\tau_s=0.036$ grains are underrepresented, even at the lower threshold of $\rho_{g,0}$, representing $\sim7\%$ of all dust mass. Meanwhile, the $\tau_s=0.191$ grains contribute just as much mass above this threshold as the larger grains.

Including observations from the dust surface density at each grain size (Fig.~\ref{fig:surfden_spec}), we can interpret the data in Figure~\ref{fig:smplot} as supporting the idea that the $\tau_s=0.191$ grains form filaments but not strong clumps, while the smaller $\tau_s=0.036$ grains form neither. In other words, the $\rho_{g,0}$ threshold appears to delimit the dust density boundary for the filamentary features.

\subsection{Simulations with larger numbers of species}
\label{sec:clump-morebins}

As with the results from Section~\ref{sec:res_surfden}, using a larger number of grain species to sample the grain size distribution does not change our results. In Table~\ref{tab:massclumps}, we include time averages of the mass in clumps and lost by clumps for the \texttt{M12} and \texttt{M18} simulations. As discussed in Section~\ref{sec:meth-moregrains}, the larger bin samples are created by sub-sampling the 6 bins from the \texttt{M6} simulations, so that we can easily combine the sub-sampled bins to match the $\tau_s$ bin boundaries from 6 bin sample for the purposes of comparison. The overall conclusions from the \texttt{M12} and \texttt{M18} data are the same: the larger $\tau_s > 0.2$ grains dominate the clump mass budget, while the dust mass lost is more evenly spread among the $\tau_s>0.1$ grains. Also, the shape of the curves from Figure~\ref{fig:smplot} are within the bounds set by the \texttt{M6} simulations.

We note that, as a whole, including Figure~\ref{fig:Cfacs}, the \texttt{M12} has slightly lower mass in clumps and dense structures than the \texttt{M6} average, while the \texttt{M18} data is slightly above this average. We do not interpret these differences as evidence that an increased number of bins affects planetesimal formation in a deterministic way. Rather, we view these differences as a consequence of the non-linear nature of the developed stage of the streaming instability. The variability in the SI is immediately observable as the range of outcomes among the individual \texttt{M6} and \texttt{S} simulations, and was the overarching theme of our previous study \citep{Rucska21}.

\subsection{Dust velocity}
\label{sec:vel-zpos}

\begin{figure*}
	\includegraphics[width=\textwidth]{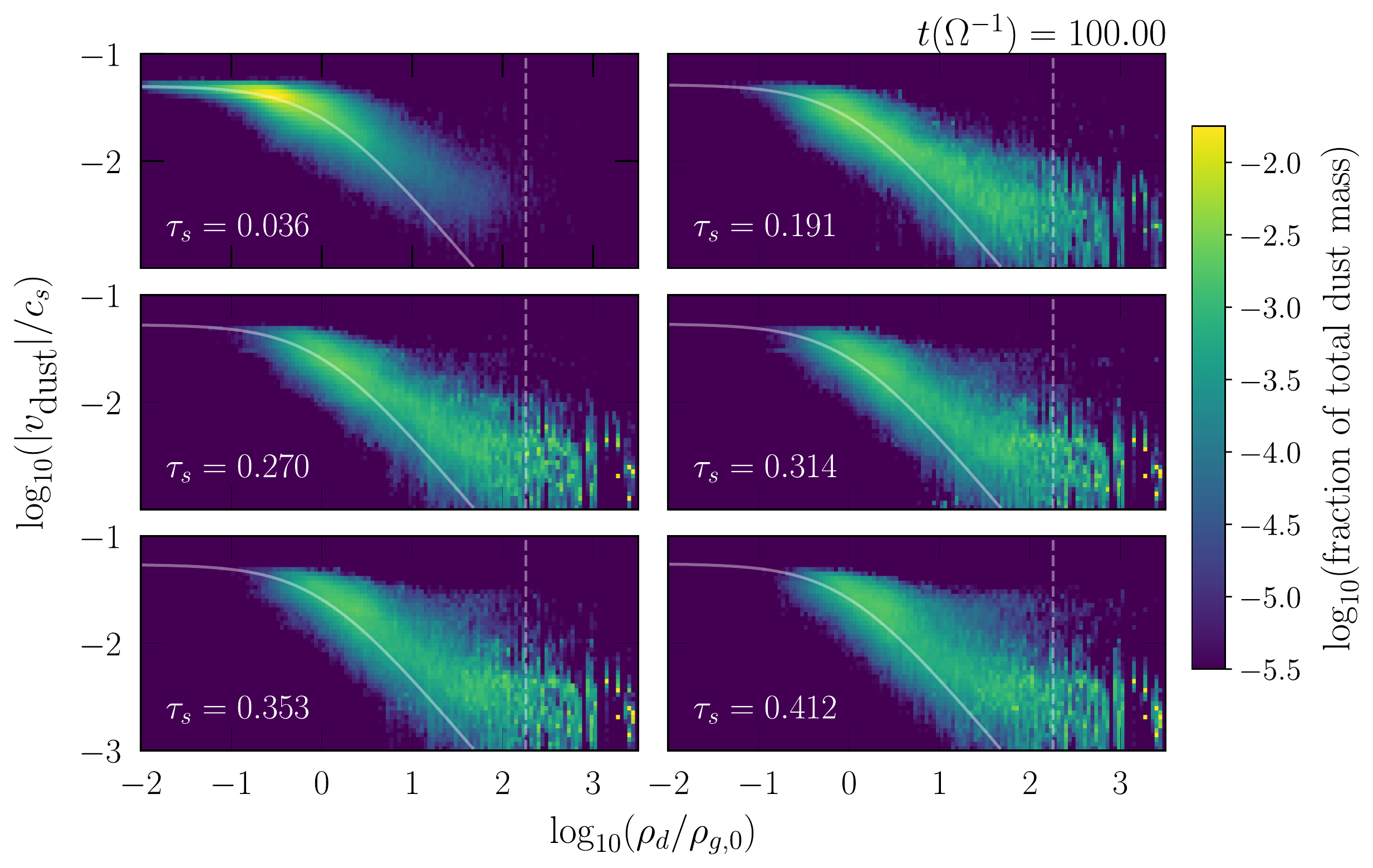}
    \caption[2D histograms in the dust density-velocity phase space]{2D histogram in the dust density-velocity phase space for the different grains in the \texttt{M6-0} simulation at $t=100\Omega^{ -1}$. Each panel is the histogram for the individual grain species. Note that all dust velocities are measured with respect to the background Keplerian flow. The (logarithmic) colourbar is normalized to the total dust mass in the simulation. The darkest bins do not contain any particles; a minimum value is applied for aesthetic purposes. The solid white curve represents the NSH equilibrium velocity (\citealt{NSH86}; see also equations 7 in \citealt{YJ07}) and the vertical white dashed line represents the Hill density in our simulation units (equation~\ref{eq:rhoH}). The NSH velocity is a function of $\tau_s$ and the local dust-to-gas mass ratio, $\epsilon=\rho_d/\rho_g$. Since $\rho_g \approx 1$ throughout our simulation domain, we use $\rho_d$ as a proxy for $\epsilon$.}
    \label{fig:dvhist}
\end{figure*}

In this section we use velocity data to further explore the differences in behavior between the smaller and larger dust grains in our simulations, and the consequences this has on planetesimal formation.

A 2D histogram of the dust particles in the dust volume density ($\rho_d$) and individual particle velocity (|$v_{\text{dust}}|$) phase space, for the \texttt{M6-0} run, is shown in Figure~\ref{fig:dvhist}. Also plotted is the magnitude of the equilibrium drift velocity for the dust \citep{NSH86} as the white curve, which tracks the expected steady-state drift rates of the dust (in the absence of complex dynamics like the non-linear SI). We see at large $\rho_d$, the expected drift velocity falls to 0, predicting that the dust fully decouples from the dust-gas equilibrium and orbits at the Keplerian velocity, and at low $\rho_d$ the drift velocity approaches to the radial pressure gradient offset $\sim \eta v_K$ with a factor of order unity that depends on $\tau_s$.

The smallest $\tau_s=0.036$ dust grains have most of their mass below $\rho_{g,0}$, which is in line with conclusions regarding Figure~\ref{fig:smplot}. Nearly all of the dust at this size--which does not form filaments or clumps--follows the NSH equilibrium curve closely. This provides further evidence that these smallest grains do not participate in highly non-linear behavior that deviates from analytical, steady-state expectations.

Most of the $\tau_s=0.191$ grains do not exist at densities above $\rho_H=180$, but between $30$ and $100\rho_{g,0}$, which corroborates earlier discussions in Section~\ref{sec:res_clumps} which conclude these grains predominantly participate in filament formation but not clump formation. The lower density dust between $\sim0.3$ and $10\rho_{g,0}$ primarily follows the NSH equilibrium curve.

For the larger $\tau_s > 0.2$ grains, most of their mass exists at large densities well above $\rho_H$. Dust in the centre of planetesimals can be seen as the bright yellow pixels at $\rho_d \ge 10 \rho_H$. The lines of above and below these brightest pixels show that the dust resolution element superparticles can have slightly different velocities within a single grid cell. As with the $\tau_s=0.191$ grains, the lower density dust is centered around the NSH expectations.

Note that for all grains with dust densities above $\sim30 \rho_{g,0}$, the bulk of the mass deviates substantially from the NSH equilibrium, settling at velocities between $\sim0.001$ and $0.01 c_s$. This is evidence of small amplitude, local turbulence, likely driven in part by the dense dust clumps near the midplane imparting substantial momentum onto the gas over small length scales. The width of the histogram about the NSH curve at lower densities is likely a result of this more disperse dust interacting with stirred up midplane gas.

\subsection{Vertical dynamics}
\label{sec:vertdyn}

\begin{figure*}
	\includegraphics[width=\textwidth]{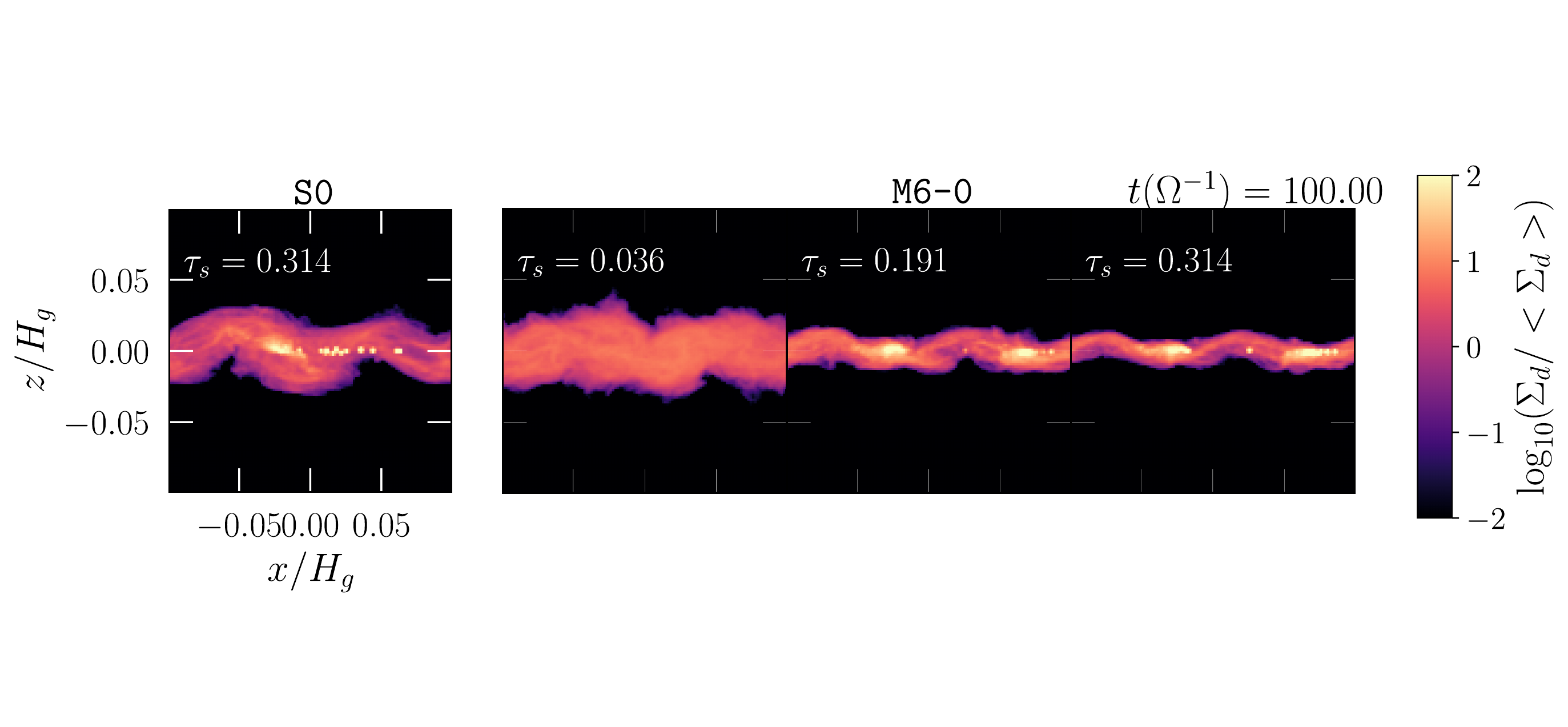}
	\caption[Dust surface density in the radial-vertical plane from the \texttt{M6-0} simulation]{Dust surface density in the $x$-$z$ (radial-vertical) plane for the \texttt{S0} simulation and a subset of grains from the \texttt{M6-0} simulation at $t=100\Omega^{ -1}$.}
    \label{fig:sdxz}
\end{figure*}

We can further highlight the different behavior between the different sized dust grains by briefly exploring the properties of the vertical (out of midplane) dynamics. Figure~\ref{fig:sdxz} shows the dust surface density in the radial-vertical ($x$-$z$) plane for the \texttt{S0} and \texttt{M6-0} simulations. In the \texttt{M6-0} run, we can see the that small grains have a much more extended vertical profile than any of the larger grains, with no bright features. Comparatively, the $\tau_s=0.314$ grains (which look nearly identical to the other grains in the largest four sizes, which are not shown) are distributed very closely to the midplane. The $\tau_s=0.191$ are slightly more extended with slightly broader features than the large grains, and the bright planetesimal between $x=0.0$ and $0.05 H_g$ is not very bright in these grains. Yet, the filament features are readily visible. In the \texttt{S0} simulation, where all grains have $\tau_s=0.314$, the vertical dust profile is much more extended than the $\tau_s=0.314$ dust in the \texttt{M6-0} run.

We can further quantify these observations by computing the dust scale height, defined as,
\begin{equation}
\label{eq:zrms}
H_p = \sqrt{\frac{1}{N_{\text{par}}} \sum_{i}^{N_{\text{par}}}
    (z_{i} - \overline{z})^2},
\end{equation}
and a similar (and related) quantity, the root mean-square (RMS) $z$ velocity,
\begin{equation}
\label{eq:vzrms}
v_{z,\text{rms}} = \sqrt{\frac{1}{N_{\text{par}}}
    \sum_{i}^{N_{\text{par}}}(v_{z,i} - \overline{v_{z}})^2}.
\end{equation}

These values for all dust grains in the \texttt{M6-0} simulation and the \texttt{S0} simulation are presented in Table~\ref{tab:vz-hp}. The $H_p$ data confirm what is visible in the vertical surface density: the smallest grains in \texttt{M6-0} have by far the most vertically extended scale heights, and the scale height monotonically decreases with $\tau_s$. The ($\tau_s=0.314$) dust in \texttt{S0} has over twice the scale height of the $\tau_s=0.314$ dust in \texttt{M6-0}.

The scale height is directly related to the RMS of the vertical dust velocity, since it is only through turbulent motions--which provide a constant source of vertical velocity dispersion--that the dust can maintain a persistent scale height \citep{YoudinLith07}. Similar to observations made by \citet{Schaffer18, Schaffer21} in their 2D simulations of the SI with multiple grains, it appears in our \texttt{M6} simulations that the larger grains stir up turbulence near the midplane, which causes the smaller grains--which are more tightly coupled to the gas aerodynamically (short drag stopping times)--to remain suspended at relatively large scale heights. The vertical RMS velocity for the gas near the midplane is $4.13\times 10^{-3}$ (in units of $c_s$), which is very close to $v_{z,\text{rms}}$ for the small $\tau_s=0.036$ grains. 
This also offers an explanation for why the $\tau_s=0.191$ dust does not form clumps, which contrasts the expectation that $\tau_s\sim 0.1$ grains are sufficiently large for planetesimal formation via the SI \citep[e.g.][]{Li21}. The larger $\tau_s>0.3$ grains generate turbulence which may impede clumping in the $\tau_s=0.191$ grains.

The \texttt{S0} dust has a larger vertical RMS velocity than any dust species in \texttt{M6-0}. Since both simulations have the same total dust mass, there is more SI-active dust in the \texttt{S0} simulation, and this leads to more turbulence and greater velocity dispersion when compared to the \texttt{M6-0} environment, where $\sim$1/6 of the dust is completely inactive in the SI, and another $\sim$1/6 does not form clumps.

\begin{table}
	\centering
	\caption[Particle scale height and vertical RMS velocity for the \texttt{M6-0} simulation]{Particle scale height and vertical RMS velocity for the different dust grains in the \texttt{M6-0} and \texttt{S0} simulations at $t=100\Omega^{ -1}$.}
	\label{tab:vz-hp}
	\begin{tabular}{lcc} % four columns, alignment for each
		\hline
		$\tau_s$ & $H_p\ (H_g)$ & $v_{z,\text{rms}}\ ( c_s )$\\
		\hline
        \rule{0pt}{3ex}\noindent
        \texttt{M6-0} & & \\
		0.036 & $11.7\times 10^{-3}$ & $4.16\times 10^{-3}$ \\
		0.191 & $4.47\times 10^{-3}$ & $2.87\times 10^{-3}$ \\
		0.270 & $3.40\times 10^{-3}$ & $2.50\times 10^{-3}$ \\
		0.314 & $3.03\times 10^{-3}$ & $2.31\times 10^{-3}$ \\
		0.353 & $2.77\times 10^{-3}$ & $2.29\times 10^{-3}$ \\
		0.412 & $2.64\times 10^{-3}$ & $2.37\times 10^{-3}$ \\
  
        \rule{0pt}{3ex}\noindent
        
        \texttt{S0} & & \\
        0.314 & $8.06\times 10^{-3}$ & $6.19\times 10^{-3}$ \\
		\hline
	\end{tabular}
\end{table}

% Discussion
\section{Conclusions and discussion}
\label{sec:discsumm3}

In this study we model a patch of a protoplanetary disc in 3D numerical hydrodynamics simulations. We model the dust component of the disc with multiple grain sizes simultaneously under conditions that are unstable to the streaming instability, and track the non-linear development of the SI to the formation of bound planetesimals. This paper extends previous work that used multiple grain sizes in simulations of the non-linear phase of the SI \citep{Johansenetal07,Bai102,Schaffer18,Schaffer21,YangZhu21}. Most prior work used a grain size distribution with a number density described by a single power law, but in our study we sample a distribution that is the output of a widely-used model of grain growth and fragmentation applicable to midplane of protoplanetary discs \citep{Birnstiel11}. To compare our multi-species results to prior work which modelled the dust with a single species, we match the peak of the size distribution to the grain size studied in \citet{Rucska21}.

Our main results are as follows:
\begin{enumerate}
    \setlength{\itemsep}{5pt}
    \item Only larger grains with dimensionless stopping times $\tau_s > 0.1$ participate strongly in the non-linear SI, producing filaments and regions with large dust densities that gravitationally collapse into planetesimals. The smaller grains do not form filaments or clumps at all, despite the fact they are embedded in an environment where roughly 5/6 of the dust mass is forming dense structures. This confirms a basic property of the multi-species SI at the non-linear stage \citep{Bai102, YangZhu21}, which remains true for a realistic protoplanetary disc grain size distribution from \citet{Birnstiel11}. The net result is there is more dust mass in the regions between the filaments in the multi-species simulations when compared to the single grain simulations, and slightly less mass in the dense structures.
    
    \item Clumping of dust via the SI on sub-AU length scales reduces the average surface brightness for a given amount of dust. This confirms in 3D models that the SI may explain the lower than expected (order unity) optical depths inferred in observed protoplanetary disc rings (see Section~\ref{sec:obscon} for details). We estimate that 20\%-80\% more dust may be present than in uniform mass distribution models. The effect is less severe for multi-size versus single-size models.  
    
     We note that there are large uncertainties regarding the dust opacity, which can depend strongly on both grain size and composition \citep{Birn18-DSV}.  Grain dependent opacities would modify the quantitative predictions from our model, but the qualitative results still apply: an active streaming instability will reduce the brightness of dust emission in protoplanetary discs due to forming denser regions (e.g. filaments) and even more so if clumping leads to planetesimals. In particular, it will selectively hide the contributions of larger grains that participate more in the streaming instability. Our results likely apply to $\sim$mm sized dust in outer disc regions that are visible to ALMA. Especially in dust rings or bumps, where the local dust-to-gas mass ratio--and the streaming instability--is enhanced, even for smaller grains.
    
    \item We identify bound clumps and dense dust features. Larger $\tau_s \gtrsim 0.2$ grains form clumps, $\tau_s \lesssim 0.04$ grains do not form clumps or filaments. Intermediate sizes are somewhat in between, forming filaments but not clumps. The velocities of the smallest grains are quite different from the larger grains in clumps and filaments, suggesting that these small grains---with a short drag stopping time that enforces tight coupling to the gas---simply sweep by the planetesimals rather than becoming incorporated into them. This implies a lower size cutoff for pebble and dust grains incorporated into asteroids and comets.
    
    \item The main group of the multi-species runs in this study used 6 bins or species to sample of the grain size distribution. We test 12 and 18 bins to show convergence. More bins appears to have no appreciable effect on the results for the multi-species simulations and we conclude that 6 bins is sufficient to study peaked grain size distributions.
\end{enumerate}

\subsection{The future of planetesimal formation via the SI with multiple grain sizes}

Including multiple sizes in models of the non-linear SI affects not just planetesimal formation but also the observable properties of protoplanetary discs. Most prior work on the SI has modelled the dust with a single grain size. However, recent observations of protoplanetary discs (see \citealt{Andrews20}, for a review) and results from grain growth theory \citep{Birnstiel11,Birnstiel15} suggest that there is a distribution of dust grain sizes within discs. An important consideration then is what the shape of this distribution should be.

In this paper we have shown that just an order of magnitude difference in grain size can determine whether grains are fully active in the SI all to way to planetesimal formation, or whether they do not even form filaments. This result motivates further exploration of the grain size distribution parameter space. Our study represents a single instance of the \citet{Birnstiel11} distribution for a specific set of disc conditions. In our results, most of the species participate in planetesimal formation. Shifting the distribution peak to smaller sizes---equivalent to considering different radial positions in the disc---would move dust mass from species that undergo strong clumping towards species that do not participate in planetesimal formation or primarily form only filaments. Presumably, this would result in an overall decrease in the total dust mass that is converted to planetesimals and may act to suppress the instability itself. Extending our work to a broader range of distributions would reveal how planetesimal formation varies in conditions at different radial locations in the disc.

Of particular interest is a distribution with a more equal mix of SI-active and SI-inactive grains. These conditions likely describe the onset of the SI and planetesimal formation. Early in the disc lifetime, most of the dust in the midplane may be too small ($\tau_s \lesssim 0.04$) to participate in planetesimal formation initially, and then grow through mutual collisions \citep[e.g.][]{Birnstiel11} to involve sizes that are unstable to the SI. However, the time scales for grain growth are typically $>10^4$ yr \citep[e.g.][]{Birnstiel12}, while the timescale for planetesimal formation via the SI is much shorter\footnote{For the timescales in our study, $100\,\Omega^{-1}\approx16$ orbital periods, which is equivalent to $\sim$200 years at 5 AU around a solar mass star.}. Thus, for initially small grains, planetesimal formation may occur as grains grow. It would be interesting to explore this initial planetesimal formation phase with a dust size distribution that includes a larger proportion of smaller, SI-inactive grains.

More realistically, however, it is likely grain growth and the streaming instability occur simultaneously. Dust growth and fragmentation is driven by collisions between dust grains. The source of the relative velocity for these collisions in models such as \citet{Birnstiel11} is an underlying turbulence that may be driven by large scale hydrodynamic instabilities \citep[see][for a review]{Lyra19}. The streaming instability generates its own turbulence locally \citep[e.g.][]{Li18} that drives relative velocities between dust, especially when a distribution of sizes is considered \citep{Bai102}. How these SI-driven collisions influence grain growth remains unstudied. A possible technique may be a model where the dust size can change based on collisions and expectations of growth/fragmentation. These dynamic grain size models have been applied to global models of disc evolution \citep[e.g.][]{Gonzalez17,Drazkowska21}, yet have not appeared in high resolution studies of the SI.

Our results show that, under the SI, a distribution of sizes will segregate spatially. The larger, pebble-sized dust settles to the midplane and undergoes vigorous non-linear dynamics leading to filament and planetesimal formation, while the smaller grains remain vertically suspended and occupy the space between filaments. Thus, the influence of grain growth likely varies spatially as well. Perhaps the small, vertically suspended grains could grow to sizes that are more SI-active, settle towards the midplane, and participate in planetesimal formation. The dense, dust-dominated regions within filaments could promote the growth of pebbles to larger sizes than is possible in gas-dominated regimes. Or, the pebbles in filaments could fragment to smaller SI-inactive grains and reduce the efficiency of planetesimal formation. These smaller sized, fragmented remnants would be created at low scale heights near the midplane, and it is unclear how those grains would interact with clumps and pebble-rich filaments. Such possibilities could be explored in dynamic grain size models.

Incorporating grain growth introduces models dependent on physical units (e.g. fragmentation threshold velocity). This breaks the scale-free property of the common shearing box model used in high-resolution studies of the SI that allows, for example, the translation of $\tau_s=0.314$ dust to represent different physical grain sizes depending on the disc model and radial position. This means multiple simulations will be required to model how grain growth theory interacts with the local dynamics of the streaming instability under different disc conditions. 

The composition of grains could also influence both grain growth and the aerodynamic coupling between the solids and gas phase. Icy grains can stick together at larger collisional velocities than silicate grains \citep[e.g.][]{Gundlach15}, and since icy grains are, generally speaking, larger than dry grains, they can radially drift through protoplanetary discs at different rates \citep{Draz17}. If both dry and icy grains co-exist in a disc region that is unstable to the SI (in the vicinity of a disc ice line), our results suggest the two populations could become spatially separated. The small, dry grains would preferentially remain suspended above the disc midplane while the larger, SI-active icy grains would form filaments and planetesimals. This would distinguish the chemical composition of the planetesimals from the overall dust population within which they are formed.

Improving numerical resolution to near planetesimal ($\sim$10 km) length scales could confirm our interpretation of our results that small grains do not participate in clump formation because they are tightly coupled to the gas which flows around the plantesimals. An increase in resolution to this scale is not possible with the methods applied to the streaming instability thus far, but may be approachable with adaptive resolution techniques and/or zoom-in simulations with small domains.

The ability of the available numerical schemes to model an aerodynamically coupled solids-gas system with a dynamic grain size distribution also remains unexplored. \citet{Bai102} suggested that 1 dust superparticle per grid cell per grain species is adequate to capture the non-linear SI, and this has been the literature standard since. It is unclear how this would translate to a dust phase with a continuous, dynamic size range. Difficulties and uncertainties aside, we believe a dynamic dust size distribution could be a promising avenue for approaching a more realistic model of planetesimal formation via the streaming instability.

\section*{Acknowledgements}

These simulations were performed on the Niagara supercomputing cluster operated by SciNet and Compute Canada. JW thanks NSERC for funding support.

\section*{Data availability}
The data underlying this article will be shared on reasonable request to the corresponding author.

%%%%%%%%%%%%%%%%%%%%%%%%%%%%%%%%%%%%%%%%%%%%%%%%%%



\bibliographystyle{mnras}
\bibliography{library} % if your bibtex file is called example.bib

%\begin{thebibliography}{99}
%\end{thebibliography}

%%%%%%%%%%%%%%%%%%%%%%%%%%%%%%%%%%%%%%%%%%%%%%%%%%

%%%%%%%%%%%%%%%%% APPENDICES %%%%%%%%%%%%%%%%%%%%%

\appendix

% Appendix section
%\input{sections/5_appendix}

%%%%%%%%%%%%%%%%%%%%%%%%%%%%%%%%%%%%%%%%%%%%%%%%%%

% Don't change these lines
\bsp	% typesetting comment
\label{lastpage}
\end{document}